\documentclass[preprint2]{emulateapj}

\usepackage{natbib}
\usepackage{longtable}
\usepackage[]{graphicx}
\usepackage{amsmath}
\usepackage{natbib}
\usepackage{tabularx}
\usepackage{bm}
\usepackage{color}
\usepackage{hyperref}

\shorttitle{Red and blue $M_{\rm BH} - M_{\rm *,sph}$ sequence}

\shortauthors{Savorgnan et al.}

\begin{document}

\title{Supermassive black holes and their host spheroids \\ II. The red and blue sequence in the $M_{\rm BH} - M_{\rm *,sph}$ diagram}

\author{Giulia A.~D.~Savorgnan and Alister W.~Graham}
\affil{Centre for Astrophysics and Supercomputing, Swinburne University of Technology, Hawthorn, Victoria 3122, Australia.}
\email{gsavorgn@astro.swin.edu.au}
\author{Alessandro Marconi}
\affil{Dipartimento di Fisica e Astronomia, Universit\'a di Firenze, via G. Sansone 1, I-50019 Sesto Fiorentino, Firenze, Italy.}
\and
\author{Eleonora Sani}
\affil{European Southern Observatory, Alonso de Cordova, Vitacura 3107, Santiago, Chile.}

\begin{abstract}
In our first paper, we performed a detailed (i.e.~bulge, disks, bars, spiral arms, rings, halo, nucleus, etc.) 
decomposition of 66 galaxies, with directly measured black hole masses, $M_{\rm BH}$, 
that had been imaged at $3.6\rm~\mu m$ with \emph{Spitzer}.
Our sample is the largest to date and, for the first time, the decompositions were checked for consistency with the galaxy kinematics. 
We present correlations between $M_{\rm BH}$ 
and the host spheroid (and galaxy) luminosity, $L_{\rm sph}$ (and $L_{\rm gal}$), 
and also stellar mass, $M_{\rm *,sph}$.
While most previous studies have used galaxy samples that were overwhelmingly dominated by high-mass, early-type galaxies,
our sample includes 17 spiral galaxies, half of which have $M_{\rm BH} < 10^7\rm~M_\odot$, 
and allows us to better investigate the poorly studied low-mass end of the $M_{\rm BH} - M_{\rm *,sph}$ correlation.
The bulges of early-type galaxies follow $M_{\rm BH} \propto M_{\rm *,sph}^{1.04 \pm 0.10}$  
and define a tight 
\emph{red sequence} with intrinsic scatter $\epsilon_{(M_{\rm BH}|M_{\rm *,sph})} = 0.43 \pm 0.06\rm~dex$ 
and a median $M_{\rm BH}/M_{\rm *,sph}$ ratio of $0.68 \pm 0.04\%$, 
i.e.~a $\pm 2\sigma$ range of 0.1--5\%.
At the low-mass end, the bulges of late-type galaxies define a much steeper 
\emph{blue sequence}, 
with $M_{\rm BH} \propto M_{\rm *,sph}^{2-3}$ and $M_{\rm BH}/M_{\rm *,sph}$ equal to $0.02\%$ at $M_{\rm BH} \approx 10^6~\rm M_\odot$, 
indicating that gas-rich processes feed the black hole more efficiently than the host bulge as they coevolve. 
We additionally report that: i) our S\'ersic galaxy sample follows $M_{\rm BH} \propto M_{\rm *,sph}^{1.48 \pm 0.20}$, 
a less steep sequence than previously reported; 
ii) bulges with S\'ersic index $n_{\rm sph}<2$, argued by some to be pseudo-bulges, 
are not offset to lower $M_{\rm BH}$ from the correlation defined by the current bulge sample with $n_{\rm sph}>2$; 
and iii) $L_{\rm sph}$ and $L_{\rm gal}$ correlate equally well with $M_{\rm BH}$, in terms of intrinsic scatter, only for early-type galaxies 
-- once reasonable numbers of spiral galaxies are included, the correlation with $L_{\rm sph}$ is better than that with $L_{\rm gal}$. 

\end{abstract}


\section{Introduction}
\label{sec:int}
A quarter of a century ago, 
\cite{dressler1989} foresaw a ``rough scaling of black hole mass with the mass of the spheroidal component'' of galaxies, 
as suggested by the sequence of five galaxies (M87, M104, M31, M32 and the Milky Way). 
\cite{yee1992} then announced a linear relation between what was effectively black hole mass and galaxy mass for high-luminosity, bulge-dominated early-type galaxies 
radiating near the Eddington limit.
This ``rough scaling'' was a premature version of the early correlations between black hole mass, $M_{\rm BH}$,  
and host spheroid luminosity, $L_{\rm sph}$, and also host spheroid mass, $M_{\rm sph}$ 
\citep{kormendyrichstone1995,magorrian1998,marconihunt2003,haringrix2004}. 
These initial studies were dominated by high-mass, early-type galaxies, 
for which they too reported a quasi-linear $M_{\rm BH} - M_{\rm sph}$ relation. 
Subsequent studies of the $M_{\rm BH} - L_{\rm sph}$ and $M_{\rm BH} - M_{\rm sph}$ diagrams 
(\citealt{ferrareseford2005,lauer2007,graham2007,gultelkin2009,sani2011,beifiori2012,erwingadotti2012,
vika2012,vandenbosch2012,mcconnellma2013,kormendyho2013})
continued to use galaxy samples dominated by high-mass, early-type systems with $M_{\rm BH} \gtrsim 0.5 \times 10^8~\rm M_\odot$, 
and they too recovered a near-linear relation. 
However, the consensus about a linear $M_{\rm BH} - M_{\rm sph}$ correlation was not unanimous. 
Some studies had reported a slope steeper than one,  
or noticed that the low-mass spheroids were offset to the right of (or below) the relation traced by the high-mass spheroids 
\citep{laor1998,wandel1999,laor2001,ryan2007}.
\cite{graham2012bent}, \cite{grahamscott2013} and \cite{scott2013} found two distinct trends in the $M_{\rm BH} - L_{\rm sph}$ and $M_{\rm BH} - M_{\rm sph}$ diagrams:  
a linear and a super-quadratic correlation at the high- and low-mass end, respectively\footnote{Readers 
interested in an extensive review about the early discovery and successive improvements of these correlations 
should consult \cite{graham2015bulges}.}. \\
Recently, \cite{lasker2014data,lasker2014anal} derived $2.2~\rm \mu m$ bulge luminosities for 35 galaxies 
(among which only 4 were classified as spiral galaxies), 
and reported a slope below unity for their $M_{\rm BH} - L_{\rm sph}$ relation. 
They also claimed that the black hole mass correlates equally well with the total galaxy luminosity 
as it does with the bulge luminosity.  \\
The $M_{\rm BH} - L_{\rm sph}$ relation for early-type (elliptical + lenticular) galaxies can be predicted by combining two other correlations that involve 
the bulge stellar velocity dispersion, $\sigma$.
One of these is the $M_{\rm BH} - \sigma$ relation \citep{ferraresemerritt2000,gebhardt2000},
which can be described with a single power-law ($M_{\rm BH} \propto \sigma^{5-6}$) 
over a wide range in velocity dispersion ($70-350~\rm km~s^{-1}$, e.g.~\citealt{graham2011,mcconnell2011,grahamscott2013}).
The other is the $L_{\rm sph} - \sigma$ relation, 
which has long been known to be a ``double power-law'', 
with $L_{\rm sph} \propto \sigma^{5-6}$ at the luminous end\footnote{Recent 
work has the $M_{\rm BH} - \sigma$ correlation as steep as $M_{\rm BH} \propto \sigma^{6.5}$ \citep{savorgnangraham2015},
and the high-luminosity end of the $L_{\rm sph} - \sigma$ correlation as steep as $L_{\rm sph} \propto \sigma^{8}$ \citep{monterodorta2015}.} 
\citep{schechter1980,malumuthkrishner1981,vonderlinden2007,lauer2007lumell,liu2008}
and $L_{\rm sph} \propto \sigma^2$ at intermediate and faint luminosities 
\citep{davies1983,held1992,matkovicguzman2005,derijcke2005,balcells2007screl,chilingarian2008,forbes2008,cody2009,tortora2009,kourkchi2012}. 
The change in slope of the $L_{\rm sph} - \sigma$ relation occurs at $M_B \approx -20.5\rm~mag$, 
corresponding to $\sigma \approx 200~\rm km~s^{-1}$. 
The $M_{\rm BH} - L_{\rm sph}$ relation should therefore be better described by a ``broken'', rather than a single power-law: 
with $M_{\rm BH} \propto L_{\rm sph}^{2.5}$ at the low-luminosity end, 
and $M_{\rm BH} \propto L_{\rm sph}^1$ at the high-luminosity end.  
Due to the scatter in the $M_{\rm BH} - L_{\rm sph}$ (or $M_{\rm BH} - M_{\rm sph}$) diagram, 
studies that have not sufficiently probed below $M_{\rm BH} \approx 10^7\rm~M_\odot$ 
can easily miss the change in slope occuring at $M_{\rm BH} \approx 10^{(8 \pm 1)}\rm~M_\odot$, 
and erroneously recover a single log-linear relation. \\
When \cite{graham2012bent} pointed out this overlooked inconsistency between these linear and bent relations, 
he identified two different populations of galaxies, 
namely the core-S\'ersic spheroids \citep{graham2003coresersicmodel,trujillo2004coresersicmodel} and the S\'ersic 
spheroids\footnote{Core-S\'ersic spheroids have partially depleted cores relative to their outer S\'ersic light profile, 
whereas S\'ersic spheroids have no central deficit of stars.},
and attributed the change in slope (from super-quadratic to linear) to their different formation mechanisms. 
In this scenario, core-S\'ersic spheroids are built in dry merger events 
where the black hole and the bulge grow at the same pace, increasing their mass in lock steps ($M_{\rm BH} \propto L_{\rm sph}^1$), 
whereas S\'ersic spheroids originate from gas-rich processes 
in which the mass of the black hole increases more rapidly than the mass of its host spheroid ($M_{\rm BH} \propto L_{\rm sph}^{2.5}$). \\
\citeauthor{grahamscott2013} (\citeyear{grahamscott2013}, hereafter GS13) and \citeauthor{scott2013} (\citeyear{scott2013}) 
presented separate power-law linear regressions 
for the S\'ersic and core-S\'ersic spheroids in the $M_{\rm BH} - L_{\rm sph}$ and $M_{\rm BH} - M_{\rm *,sph}$ 
(spheroid stellar mass) diagrams, probing down to $M_{\rm BH} \approx 10^6\rm~M_\odot$. 
To obtain their dust-corrected \emph{bulge} magnitudes, they did not perform bulge/disc decompositions, 
but converted the $B-$band and $K_S-$band observed, total \emph{galaxy} magnitudes into bulge magnitudes 
using a mean statistical bulge-to-total ratio based on each object's morphological type and disc 
inclination\footnote{While this resulted in individual bulge magnitudes not being exactly correct, 
their large sample size allowed them to obtain a reasonable $M_{\rm BH} - L_{\rm sph}$ relation for the ensemble.}. 
These mean statistical bulge-to-total ratios were obtained from the results of two-component (S\'ersic-bulge/exponential-disk) decompositions in the literature. \\
Here we investigate in more detail 
the $M_{\rm BH} - L_{\rm sph}$ and $M_{\rm BH} - M_{\rm *,sph}$ diagrams 
using state-of-the-art galaxy decompositions (Savorgnan \& Graham 2015, hereafter \emph{Paper I}) 
for galaxies with directly measured black hole masses.
Our galaxies are large and nearby, which allows us to perform accurate multicomponent decompositions 
(instead of simple bulge/disk decompositions). 
Our decompositions were performed on $3.6\rm~\mu m$ \emph{Spitzer} satellite imagery, 
which is an excellent proxy for the stellar mass, superior to the $K-$band (\citealt{sheth2010}, and references therein).
Nine of our galaxies have $M_{\rm BH} \lesssim 10^7\rm~M_\odot$, 
which allows us to better constrain the slope of the correlation at the low-mass end.
Furthermore, our galaxy sample includes 17 spiral galaxies, 
representing a notable improvement over past studies dominated by early-type galaxies. 
In a forthcoming paper, we will explore the relation between the black hole mass and the bulge dynamical mass, 
$M_{\rm dyn,sph} \propto R_{\rm e} \sigma^2$, and address the issue of a black hole fundamental plane.

\section{Data}
\label{sec:data}
Our galaxy sample (see Table \ref{tab:sample}) 
consists of 66 objects for which a dynamical measurement of the black hole mass had been tabulated in the literature 
(by GS13 or \citealt{rusli2013bhmassesDM}) at the time that we started this project, 
and for which we were able to obtain useful spheroid parameters from $3.6\rm~\mu m$ \emph{Spitzer} satellite imagery. \\
Spheroid magnitudes were derived from our state-of-the-art galaxy decompositions, which take into account 
bulge, disks, spiral arms, bars, rings, halo, extended or unresolved nuclear source and partially depleted core. 
Kinematical information \citep{atlas3dIII,scott2014,arnold2014} was used 
to confirm the presence of rotationally supported disk components in most early-type (elliptical + lenticular) galaxies, 
and to identify their extent 
(intermediate-scale disks that are fully embedded in the bulge, 
or large-scale disks that encase the bulge and dominate the light at large radii). 
It is worth stressing that, contrary to common knowledge, the majority of ``elliptical'' galaxies contain disks,
i.e.~they are not single-component spheroidal systems.
\emph{Paper I} presents the dataset used here, 
including details about the data reduction process and the galaxy modelling technique that we developed. 
It also discusses how we estimated the uncertainties\footnote{By comparing, for each of our galaxies, the measurements of the bulge magnitude 
obtained by different authors with that obtained by us, we estimated the uncertainties on the bulge magnitudes, 
in effect taking into account systematic errors. 
Systematic errors include incorrect sky subtraction, inaccurate masking of contaminating sources, imprecise description of the PSF, 
erroneous choice of model components (for example, when failing to identify a galaxy subcomponent and thus omitting it in the model, 
or when describing a galaxy sub-component with an inadequate function), 
the radial extent of the surface brightness profile and one's sampling of this. 
Most of these factors are not included in popular 2D fitting codes which report only the statistical errors associated with their fitted parameters. 
In fact, when performing multi-component decomposition of high signal-to-noise images of nearby -- therefore well spatially resolved -- galaxies, 
errors are dominated by systematics rather than Poisson noise. 
Unlike many papers, we believe that we have not under-estimated the uncertainties associated to the bulge best-fit parameters. } 
on the bulge magnitudes, and presents the individual 66 galaxy decompositions, 
along with a comparison and discussion of past decompositions. \\
Bulge luminosities\footnote{Following \cite{sani2011}, absolute luminosities were calculated 
assuming a $3.6\rm~\mu m$ solar absolute magnitude of $3.25\rm~mag$. 
Absolute luminosities were not corrected for cosmological redshift dimming 
(this correction would be as small as $-0.02\rm~mag$ for galaxies at a distance of $40\rm~Mpc$ 
or $-0.05\rm~mag$ for galaxies at a distance of $100\rm~Mpc$).} 
(Table \ref{tab:sample}) from \emph{Paper I} were converted into stellar masses 
using a constant $3.6\rm~\mu m$ mass-to-light ratio, $\Gamma_{3.6} = 0.6$ \citep{meidt2014}.
We additionally explored a more sophisticated way to compute mass-to-light ratios, 
using the color-$\Gamma_{3.6}$ relation published by 
\citeauthor{meidt2014} (\citeyear{meidt2014}, their equation 4), 
which allows one to estimate $\Gamma_{3.6}$ of a galaxy from its $[3.6] - [4.5]$ color. 
Individual $[3.6] - [4.5]$ colors\footnote{These are integrated $[3.6] - [4.5]$ colors, measured in a circular aperture 
within each galaxy's effective radius.} were taken from 
\citeauthor{peletier2012} (\citeyear{peletier2012}, column 8 of their Table 1) 
when available for our galaxies, 
or were estimated from the bulge stellar velocity dispersion, $\sigma$, 
using the color-$\sigma$ relation presented by \citeauthor{peletier2012} (\citeyear{peletier2012}, their Figure 6).
We found that the range in $[3.6] - [4.5]$ color is small ($0.06\rm~mag$), 
and thus the range in $\Gamma_{3.6}$ is also small ($0.04$).
After checking that using a single $\Gamma_{3.6} = 0.6$, independent of $[3.6] - [4.5]$ color, 
does not significantly affect the results of our analysis, 
we decided to use individual, color-dependent mass-to-light ratios. \\
For each galaxy, the total luminosity (or galaxy luminosity, $L_{\rm gal}$) is the sum of the luminosities of all its sub-components. 
Due to the complexity of their modelling, 
four galaxies (see Table \ref{tab:sample}, column 7) had their galaxy luminosities 
underestimated\footnote{These four cases are discussed in \emph{Paper I}.}, 
which are given here as lower limits. 
Following GS13, we assumed a fixed uncertainty ($0.25\rm~mag$) for the absolute galaxy magnitude $MAG_{\rm gal}$. \\
The morphological classification (E = elliptical; E/S0 = elliptical/lenticular; S0 = lenticular; S0/Sp = lenticular/spiral; Sp = spiral; and ``merger'') 
follows from the galaxy models presented in \emph{Paper I}. 
Throughout this paper we will refer to early-type galaxies (E+S0) and late-type galaxies (Sp). 
Two galaxies classified as E/S0 are obviously included in the early-type bin, 
whereas two galaxies classified as S0/Sp and another two classified as mergers are included in neither the early- nor the late-type bin.\\
The S\'ersic/core-S\'ersic classification presented in this work 
comes from the compilation of \citet{savorgnangraham2015},
who identified partially depleted cores according to the same criteria used by GS13.
When no high-resolution image analysis was available from the literature, 
they inferred the presence of a partially depleted core based on the stellar velocity dispersion:
a spheroid is classified as core-S\'ersic if $\sigma > 270\rm~km~s^{-1}$,
or as S\'ersic if $\sigma < 166\rm~km~s^{-1}$. 
All of the galaxies with velocity dispersions between these two limits had high-resolution images available. 

\begin{table*}                                        
\small                                                
\begin{center}                                        
\caption{Galaxy sample.} 
\begin{tabular}{llllllrll}                           
\tableline                                                
\multicolumn{1}{l}{{\bf Galaxy}} &                   
\multicolumn{1}{l}{{\bf Type}} &                     
\multicolumn{1}{l}{{\bf Core}} &                     
\multicolumn{1}{l}{{\bf Distance}} &                 
\multicolumn{1}{l}{{\bf $\bm{M_{\rm BH}}$}} &  
\multicolumn{1}{l}{{\bf $\bm{MAG_{\rm sph}}$}} &  
\multicolumn{1}{l}{{\bf $\bm{MAG_{\rm gal}}$}} &  
\multicolumn{1}{l}{{\bf $\bm{[3.6]-[4.5]}$}} &  
\multicolumn{1}{l}{{\bf $\bm{M_{\rm *,sph}}$}} \\  
\multicolumn{1}{l}{} &                                
\multicolumn{1}{l}{} &                                
\multicolumn{1}{l}{} &                                
\multicolumn{1}{l}{[Mpc]} &                           
\multicolumn{1}{l}{$[10^8~\rm M_{\odot}]$} &         
\multicolumn{1}{l}{[mag]} &                                
\multicolumn{1}{l}{[mag]} &                                
\multicolumn{1}{l}{[mag]} &                                
\multicolumn{1}{l}{$[10^{10}~\rm M_{\odot}]$} \\                             
\multicolumn{1}{l}{(1)} &                             
\multicolumn{1}{l}{(2)} &                             
\multicolumn{1}{l}{(3)} &                             
\multicolumn{1}{l}{(4)} &                             
\multicolumn{1}{l}{(5)} &                             
\multicolumn{1}{l}{(6)} &                             
\multicolumn{1}{l}{(7)} &                             
\multicolumn{1}{l}{(8)} &                             
\multicolumn{1}{l}{(9)} \\                         
\tableline                                                
IC 1459  &  E  &  yes   &  $28.4$  &  $24_{-10}^{+10}$   &  $-26.15_{-0.11}^{+0.18}$   &  $-26.15 \pm 0.25$ 
 &  $-0.12$  &  $27_{-4}^{+3}$   \\ 
IC 2560  &  Sp (bar)  &  no?  &  $40.7$  &  $0.044_{-0.022}^{+0.044}$   &  $-22.27_{-0.58}^{+0.66}$   &  $-24.76 \pm 0.25$ 
 &  $-0.08$  &  $1.0_{-0.5}^{+0.7}$   \\ 
IC 4296  &  E  &  yes?  &  $40.7$  &  $11_{-2}^{+2}$   &  $-26.35_{-0.11}^{+0.18}$   &  $-26.35 \pm 0.25$ 
 &  $-0.12$  &  $31_{-5}^{+3}$   \\ 
M31  &  Sp (bar)  &  no   &  $0.7$  &  $1.4_{-0.3}^{+0.9}$   &  $-22.74_{-0.11}^{+0.18}$   &  $-24.67 \pm 0.25$ 
 &  $-0.09$  &  $1.5_{-0.2}^{+0.2}$   \\ 
M49  &  E  &  yes   &  $17.1$  &  $25_{-1}^{+3}$   &  $-26.54_{-0.11}^{+0.18}$   &  $-26.54 \pm 0.25$ 
 &  $-0.12$  &  $39_{-6}^{+4}$   \\ 
M59  &  E  &  no   &  $17.8$  &  $3.9_{-0.4}^{+0.4}$   &  $-25.18_{-0.11}^{+0.18}$   &  $-25.27 \pm 0.25$ 
 &  $-0.09$  &  $14_{-2}^{+2}$   \\ 
M64  &  Sp  &  no?  &  $7.3$  &  $0.016_{-0.004}^{+0.004}$   &  $-21.54_{-0.11}^{+0.18}$   &  $-24.24 \pm 0.25$ 
 &  $-0.06$  &  $0.64_{-0.10}^{+0.07}$   \\ 
M81  &  Sp (bar)  &  no   &  $3.8$  &  $0.74_{-0.11}^{+0.21}$   &  $-23.01_{-0.66}^{+0.88}$   &  $-24.43 \pm 0.25$ 
 &  $-0.09$  &  $1.9_{-1.1}^{+1.6}$   \\ 
M84  &  E  &  yes   &  $17.9$  &  $9.0_{-0.8}^{+0.9}$   &  $-26.01_{-0.58}^{+0.66}$   &  $-26.01 \pm 0.25$ 
 &  $-0.10$  &  $28_{-13}^{+20}$   \\ 
M87  &  E  &  yes   &  $15.6$  &  $58.0_{-3.5}^{+3.5}$   &  $-26.00_{-0.58}^{+0.66}$   &  $-26.00 \pm 0.25$ 
 &  $-0.11$  &  $26_{-12}^{+18}$   \\ 
M89  &  E  &  yes   &  $14.9$  &  $4.7_{-0.5}^{+0.5}$   &  $-24.48_{-0.58}^{+0.66}$   &  $-24.74 \pm 0.25$ 
 &  $-0.11$  &  $6.3_{-2.9}^{+4.4}$   \\ 
M94  &  Sp (bar)  &  no?  &  $4.4$  &  $0.060_{-0.014}^{+0.014}$   &  $-22.08_{-0.11}^{+0.18}$   &  $\leq-23.36$   &  $-0.07$  &  $1.00_{-0.15}^{+0.11}$   \\ 
M96  &  Sp (bar)  &  no   &  $10.1$  &  $0.073_{-0.015}^{+0.015}$   &  $-22.15_{-0.11}^{+0.18}$   &  $-24.20 \pm 0.25$ 
 &  $-0.08$  &  $0.97_{-0.15}^{+0.11}$   \\ 
M104  &  S0/Sp  &  yes   &  $9.5$  &  $6.4_{-0.4}^{+0.4}$   &  $-23.91_{-0.58}^{+0.66}$   &  $-25.21 \pm 0.25$ 
 &  $-0.12$  &  $3.4_{-1.6}^{+2.4}$   \\ 
M105  &  E  &  yes   &  $10.3$  &  $4_{-1}^{+1}$   &  $-24.29_{-0.58}^{+0.66}$   &  $-24.29 \pm 0.25$ 
 &  $-0.10$  &  $5.6_{-2.5}^{+3.9}$   \\ 
M106  &  Sp (bar)  &  no   &  $7.2$  &  $0.39_{-0.01}^{+0.01}$   &  $-21.11_{-0.11}^{+0.18}$   &  $-24.04 \pm 0.25$ 
 &  $-0.08$  &  $0.37_{-0.06}^{+0.04}$   \\ 
NGC 0524  &  S0  &  yes   &  $23.3$  &  $8.3_{-1.3}^{+2.7}$   &  $-23.19_{-0.11}^{+0.18}$   &  $-24.92 \pm 0.25$ 
 &  $-0.09$  &  $2.2_{-0.3}^{+0.2}$   \\ 
NGC 0821  &  E  &  no   &  $23.4$  &  $0.39_{-0.09}^{+0.26}$   &  $-24.00_{-0.66}^{+0.88}$   &  $-24.26 \pm 0.25$ 
 &  $-0.09$  &  $4.7_{-2.6}^{+4.0}$   \\ 
NGC 1023  &  S0 (bar)  &  no   &  $11.1$  &  $0.42_{-0.04}^{+0.04}$   &  $-22.82_{-0.11}^{+0.18}$   &  $-24.20 \pm 0.25$ 
 &  $-0.10$  &  $1.5_{-0.2}^{+0.2}$   \\ 
NGC 1300  &  Sp (bar)  &  no   &  $20.7$  &  $0.73_{-0.35}^{+0.69}$   &  $-22.06_{-0.58}^{+0.66}$   &  $-24.16 \pm 0.25$ 
 &  $-0.10$  &  $0.70_{-0.32}^{+0.49}$   \\ 
NGC 1316  &  merger  &  no   &  $18.6$  &  $1.50_{-0.80}^{+0.75}$   &  $-24.89_{-0.58}^{+0.66}$   &  $-26.48 \pm 0.25$ 
 &  $-0.10$  &  $9.5_{-4.3}^{+6.7}$   \\ 
NGC 1332  &  E/S0  &  no   &  $22.3$  &  $14_{-2}^{+2}$   &  $-24.89_{-0.66}^{+0.88}$   &  $-24.95 \pm 0.25$ 
 &  $-0.12$  &  $8.2_{-4.5}^{+6.8}$   \\ 
NGC 1374  &  E  &  no?  &  $19.2$  &  $5.8_{-0.5}^{+0.5}$   &  $-23.68_{-0.11}^{+0.18}$   &  $-23.70 \pm 0.25$ 
 &  $-0.09$  &  $3.6_{-0.5}^{+0.4}$   \\ 
NGC 1399  &  E  &  yes   &  $19.4$  &  $4.7_{-0.6}^{+0.6}$   &  $-26.43_{-0.11}^{+0.18}$   &  $-26.46 \pm 0.25$ 
 &  $-0.12$  &  $33_{-5}^{+4}$   \\ 
NGC 2273  &  Sp (bar)  &  no   &  $28.5$  &  $0.083_{-0.004}^{+0.004}$   &  $-23.00_{-0.58}^{+0.66}$   &  $-24.21 \pm 0.25$ 
 &  $-0.08$  &  $2.0_{-0.9}^{+1.4}$   \\ 
NGC 2549  &  S0 (bar)  &  no   &  $12.3$  &  $0.14_{-0.13}^{+0.02}$   &  $-21.25_{-0.11}^{+0.18}$   &  $-22.60 \pm 0.25$ 
 &  $-0.10$  &  $0.35_{-0.05}^{+0.04}$   \\ 
NGC 2778  &  S0 (bar)  &  no   &  $22.3$  &  $0.15_{-0.10}^{+0.09}$   &  $-20.80_{-0.58}^{+0.66}$   &  $-22.44 \pm 0.25$ 
 &  $-0.09$  &  $0.25_{-0.12}^{+0.18}$   \\ 
NGC 2787  &  S0 (bar)  &  no   &  $7.3$  &  $0.40_{-0.05}^{+0.04}$   &  $-20.11_{-0.58}^{+0.66}$   &  $-22.28 \pm 0.25$ 
 &  $-0.10$  &  $0.12_{-0.05}^{+0.08}$   \\ 
NGC 2974  &  Sp (bar)  &  no   &  $20.9$  &  $1.7_{-0.2}^{+0.2}$   &  $-22.95_{-0.58}^{+0.66}$   &  $-24.16 \pm 0.25$ 
 &  $-0.09$  &  $1.8_{-0.8}^{+1.3}$   \\ 
NGC 3079  &  Sp (bar)  &  no?  &  $20.7$  &  $0.024_{-0.012}^{+0.024}$   &  $-23.01_{-0.58}^{+0.66}$   &  $\leq-24.45$   &  $-0.07$  &  $2.4_{-1.1}^{+1.7}$   \\ 
NGC 3091  &  E  &  yes   &  $51.2$  &  $36_{-2}^{+1}$   &  $-26.28_{-0.11}^{+0.18}$   &  $-26.28 \pm 0.25$ 
 &  $-0.12$  &  $30_{-5}^{+3}$   \\ 
NGC 3115  &  E/S0  &  no   &  $9.4$  &  $8.8_{-2.7}^{+10.0}$   &  $-24.22_{-0.11}^{+0.18}$   &  $-24.40 \pm 0.25$ 
 &  $-0.11$  &  $4.9_{-0.7}^{+0.5}$   \\ 
NGC 3227  &  Sp (bar)  &  no   &  $20.3$  &  $0.14_{-0.06}^{+0.10}$   &  $-21.76_{-0.58}^{+0.66}$   &  $-24.26 \pm 0.25$ 
 &  $-0.08$  &  $0.67_{-0.31}^{+0.47}$   \\ 
NGC 3245  &  S0 (bar)  &  no   &  $20.3$  &  $2.0_{-0.5}^{+0.5}$   &  $-22.43_{-0.11}^{+0.18}$   &  $-23.88 \pm 0.25$ 
 &  $-0.10$  &  $1.0_{-0.2}^{+0.1}$   \\ 
NGC 3377  &  E  &  no   &  $10.9$  &  $0.77_{-0.06}^{+0.04}$   &  $-23.49_{-0.58}^{+0.66}$   &  $-23.57 \pm 0.25$ 
 &  $-0.06$  &  $4.0_{-1.8}^{+2.8}$   \\ 
NGC 3384  &  S0 (bar)  &  no   &  $11.3$  &  $0.17_{-0.02}^{+0.01}$   &  $-22.43_{-0.11}^{+0.18}$   &  $-23.74 \pm 0.25$ 
 &  $-0.08$  &  $1.2_{-0.2}^{+0.1}$   \\ 
NGC 3393  &  Sp (bar)  &  no   &  $55.2$  &  $0.34_{-0.02}^{+0.02}$   &  $-23.48_{-0.58}^{+0.66}$   &  $-25.29 \pm 0.25$ 
 &  $-0.10$  &  $2.8_{-1.3}^{+1.9}$   \\ 
NGC 3414  &  E  &  no   &  $24.5$  &  $2.4_{-0.3}^{+0.3}$   &  $-24.35_{-0.11}^{+0.18}$   &  $-24.42 \pm 0.25$ 
 &  $-0.09$  &  $6.5_{-1.0}^{+0.7}$   \\ 
NGC 3489  &  S0/Sp (bar)  &  no   &  $11.7$  &  $0.058_{-0.008}^{+0.008}$   &  $-21.13_{-0.58}^{+0.66}$   &  $-23.07 \pm 0.25$ 
 &  $-0.06$  &  $0.42_{-0.19}^{+0.30}$   \\ 
NGC 3585  &  E  &  no   &  $19.5$  &  $3.1_{-0.6}^{+1.4}$   &  $-25.52_{-0.58}^{+0.66}$   &  $-25.55 \pm 0.25$ 
 &  $-0.10$  &  $18_{-8}^{+12}$   \\ 
NGC 3607  &  E  &  no   &  $22.2$  &  $1.3_{-0.5}^{+0.5}$   &  $-25.36_{-0.58}^{+0.66}$   &  $-25.45 \pm 0.25$ 
 &  $-0.10$  &  $15_{-7}^{+10}$   \\ 
NGC 3608  &  E  &  yes   &  $22.3$  &  $2.0_{-0.6}^{+1.1}$   &  $-24.50_{-0.58}^{+0.66}$   &  $-24.50 \pm 0.25$ 
 &  $-0.08$  &  $7.8_{-3.6}^{+5.5}$   \\ 
NGC 3842  &  E  &  yes   &  $98.4$  &  $97_{-26}^{+30}$   &  $-27.00_{-0.11}^{+0.18}$   &  $-27.04 \pm 0.25$ 
 &  $-0.11$  &  $61_{-9}^{+7}$   \\ 
NGC 3998  &  S0 (bar)  &  no   &  $13.7$  &  $8.1_{-1.9}^{+2.0}$   &  $-22.32_{-0.66}^{+0.88}$   &  $-23.53 \pm 0.25$ 
 &  $-0.12$  &  $0.78_{-0.43}^{+0.65}$   \\ 
NGC 4026  &  S0 (bar)  &  no   &  $13.2$  &  $1.8_{-0.3}^{+0.6}$   &  $-21.58_{-0.66}^{+0.88}$   &  $-23.16 \pm 0.25$ 
 &  $-0.09$  &  $0.50_{-0.28}^{+0.42}$   \\ 
NGC 4151  &  Sp (bar)  &  no   &  $20.0$  &  $0.65_{-0.07}^{+0.07}$   &  $-23.40_{-0.58}^{+0.66}$   &  $-24.44 \pm 0.25$ 
 &  $-0.09$  &  $2.8_{-1.3}^{+2.0}$   \\ 
NGC 4261  &  E  &  yes   &  $30.8$  &  $5_{-1}^{+1}$   &  $-25.72_{-0.58}^{+0.66}$   &  $-25.76 \pm 0.25$ 
 &  $-0.12$  &  $18_{-8}^{+13}$   \\ 
NGC 4291  &  E  &  yes   &  $25.5$  &  $3.3_{-2.5}^{+0.9}$   &  $-24.05_{-0.58}^{+0.66}$   &  $-24.05 \pm 0.25$ 
 &  $-0.11$  &  $3.9_{-1.8}^{+2.8}$   \\ 
NGC 4388  &  Sp (bar)  &  no?  &  $17.0$  &  $0.075_{-0.002}^{+0.002}$   &  $-21.26_{-0.66}^{+0.88}$   &  $\leq-23.50$   &  $-0.07$  &  $0.46_{-0.26}^{+0.39}$   \\ 
NGC 4459  &  S0  &  no   &  $15.7$  &  $0.68_{-0.13}^{+0.13}$   &  $-23.48_{-0.58}^{+0.66}$   &  $-24.01 \pm 0.25$ 
 &  $-0.09$  &  $2.9_{-1.3}^{+2.1}$   \\ 
NGC 4473  &  E  &  no   &  $15.3$  &  $1.2_{-0.9}^{+0.4}$   &  $-23.88_{-0.58}^{+0.66}$   &  $-24.11 \pm 0.25$ 
 &  $-0.10$  &  $3.9_{-1.8}^{+2.7}$   \\ 
NGC 4564  &  S0  &  no   &  $14.6$  &  $0.60_{-0.09}^{+0.03}$   &  $-22.30_{-0.11}^{+0.18}$   &  $-22.99 \pm 0.25$ 
 &  $-0.11$  &  $0.82_{-0.12}^{+0.09}$   \\ 
NGC 4596  &  S0 (bar)  &  no   &  $17.0$  &  $0.79_{-0.33}^{+0.38}$   &  $-22.73_{-0.11}^{+0.18}$   &  $-24.18 \pm 0.25$ 
 &  $-0.08$  &  $1.6_{-0.2}^{+0.2}$   \\ 
\tableline         
\end{tabular}   
\label{tab:sample} 
\end{center}    
\end{table*}    

\begin{table*}                                        
\small                                                
\begin{center}                                        
\begin{tabular}{llllllrll}                           
\tableline                                                
\multicolumn{1}{l}{{\bf Galaxy}} &                   
\multicolumn{1}{l}{{\bf Type}} &                     
\multicolumn{1}{l}{{\bf Core}} &                     
\multicolumn{1}{l}{{\bf Distance}} &                 
\multicolumn{1}{l}{{\bf $\bm{M_{\rm BH}}$}} &  
\multicolumn{1}{l}{{\bf $\bm{MAG_{\rm sph}}$}} &  
\multicolumn{1}{l}{{\bf $\bm{MAG_{\rm gal}}$}} &  
\multicolumn{1}{l}{{\bf $\bm{[3.6]-[4.5]}$}} &  
\multicolumn{1}{l}{{\bf $\bm{M_{\rm *,sph}}$}} \\  
\multicolumn{1}{l}{} &                                
\multicolumn{1}{l}{} &                                
\multicolumn{1}{l}{} &                                
\multicolumn{1}{l}{[Mpc]} &                           
\multicolumn{1}{l}{$[10^8~\rm M_{\odot}]$} &         
\multicolumn{1}{l}{[mag]} &                                
\multicolumn{1}{l}{[mag]} &                                
\multicolumn{1}{l}{[mag]} &                                
\multicolumn{1}{l}{$[10^{10}~\rm M_{\odot}]$} \\                             
\multicolumn{1}{l}{(1)} &                             
\multicolumn{1}{l}{(2)} &                             
\multicolumn{1}{l}{(3)} &                             
\multicolumn{1}{l}{(4)} &                             
\multicolumn{1}{l}{(5)} &                             
\multicolumn{1}{l}{(6)} &                             
\multicolumn{1}{l}{(7)} &                             
\multicolumn{1}{l}{(8)} &                             
\multicolumn{1}{l}{(9)} \\                         
\tableline                                                
NGC 4697  &  E  &  no   &  $11.4$  &  $1.8_{-0.1}^{+0.2}$   &  $-24.82_{-0.66}^{+0.88}$   &  $-24.94 \pm 0.25$ 
 &  $-0.09$  &  $10_{-6}^{+8}$   \\ 
NGC 4889  &  E  &  yes   &  $103.2$  &  $210_{-160}^{+160}$   &  $-27.54_{-0.11}^{+0.18}$   &  $-27.54 \pm 0.25$ 
 &  $-0.12$  &  $91_{-14}^{+10}$   \\ 
NGC 4945  &  Sp (bar)  &  no?  &  $3.8$  &  $0.014_{-0.007}^{+0.014}$   &  $-20.96_{-0.58}^{+0.66}$   &  $\leq-23.79$   &  $-0.06$  &  $0.36_{-0.17}^{+0.26}$   \\ 
NGC 5077  &  E  &  yes   &  $41.2$  &  $7.4_{-3.0}^{+4.7}$   &  $-25.45_{-0.11}^{+0.18}$   &  $-25.45 \pm 0.25$ 
 &  $-0.11$  &  $15_{-2}^{+2}$   \\ 
NGC 5128  &  merger  &  no?  &  $3.8$  &  $0.45_{-0.10}^{+0.17}$   &  $-23.89_{-0.66}^{+0.88}$   &  $-24.97 \pm 0.25$ 
 &  $-0.07$  &  $5.0_{-2.8}^{+4.2}$   \\ 
NGC 5576  &  E  &  no   &  $24.8$  &  $1.6_{-0.4}^{+0.3}$   &  $-24.44_{-0.11}^{+0.18}$   &  $-24.44 \pm 0.25$ 
 &  $-0.09$  &  $7.1_{-1.1}^{+0.8}$   \\ 
NGC 5845  &  S0  &  no   &  $25.2$  &  $2.6_{-1.5}^{+0.4}$   &  $-22.96_{-0.66}^{+0.88}$   &  $-23.10 \pm 0.25$ 
 &  $-0.12$  &  $1.4_{-0.8}^{+1.2}$   \\ 
NGC 5846  &  E  &  yes   &  $24.2$  &  $11_{-1}^{+1}$   &  $-25.81_{-0.58}^{+0.66}$   &  $-25.81 \pm 0.25$ 
 &  $-0.10$  &  $22_{-10}^{+16}$   \\ 
NGC 6251  &  E  &  yes?  &  $104.6$  &  $5_{-2}^{+2}$   &  $-26.75_{-0.11}^{+0.18}$   &  $-26.75 \pm 0.25$ 
 &  $-0.12$  &  $46_{-7}^{+5}$   \\ 
NGC 7052  &  E  &  yes   &  $66.4$  &  $3.7_{-1.5}^{+2.6}$   &  $-26.32_{-0.11}^{+0.18}$   &  $-26.32 \pm 0.25$ 
 &  $-0.11$  &  $33_{-5}^{+4}$   \\ 
NGC 7619  &  E  &  yes   &  $51.5$  &  $25_{-3}^{+8}$   &  $-26.35_{-0.58}^{+0.66}$   &  $-26.41 \pm 0.25$ 
 &  $-0.11$  &  $33_{-15}^{+23}$   \\ 
NGC 7768  &  E  &  yes   &  $112.8$  &  $13_{-4}^{+5}$   &  $-26.90_{-0.58}^{+0.66}$   &  $-26.90 \pm 0.25$ 
 &  $-0.11$  &  $57_{-26}^{+40}$   \\ 
UGC 03789  &  Sp (bar)  &  no?  &  $48.4$  &  $0.108_{-0.005}^{+0.005}$   &  $-22.77_{-0.66}^{+0.88}$   &  $-24.20 \pm 0.25$ 
 &  $-0.07$  &  $1.9_{-1.0}^{+1.6}$   \\ 
\tableline         
\end{tabular}   
\tablecomments{\emph{Column (1)}: Galaxy name. 
\emph{Column (2)}: Morphological type (E=elliptical, S0=lenticular, Sp=spiral, merger). 		       The morphological classification of four galaxies is uncertain (E/S0 or S0/Sp). 		       The presence of a bar is indicated. 
\emph{Column (3)}: Presence of a partially depleted core. 			The question mark is used when the classification has come from the velocity dispersion criteria mentioned in Section \ref{sec:data}. 
\emph{Column (4)}: Distance. 
\emph{Column (5)}: Black hole mass. 
\emph{Column (6)}: Absolute $3.6\rm~\mu m$ bulge magnitude. 		       Bulge magnitudes come from our state-of-the-art multicomponent galaxy decompositions (\emph{Paper I}), 		       which include bulges, disks, bars, spiral arms, rings, haloes, extended or unresolved nuclear sources and partially depleted cores,                        and that -- for the first time -- were checked to be consistent with the galaxy kinematics. 		       The uncertainties were estimated with a method that takes into account systematic errors, which are typically not considered by popular 2D fitting codes. 
\emph{Column (7)}: Absolute $3.6\rm~\mu m$ galaxy magnitude. 			Four galaxies had their magnitudes overestimated, which are given here as upper limits. 
\emph{Column (8)}: $[3.6]-[4.5]$ colour. 
\emph{Column (9)}: Bulge stellar mass. } 
\end{center}    
\end{table*}

\section{Analysis}
\label{sec:anal}
We performed a linear regression analysis of the $M_{\rm BH} - L_{\rm gal}$ (see Table \ref{tab:lreggal}), 
$M_{\rm BH} - L_{\rm sph}$ (see Table \ref{tab:lregsph}) and $M_{\rm BH} - M_{\rm *,sph}$ (see Table \ref{tab:lregmass}) data,
using the BCES code from \cite{akritasbershady1996}. 
We also repeated the analysis using both the FITEXY routine \citep{press1992}, as modified by \cite{tremaine2002}, 
and the Bayesian estimator {\tt linmix\_err} \citep{linmixerr}. 
All of these three linear regression routines account for the intrinsic scatter, 
but only the last two allow one to quantify it.
We report linear regressions, both symmetrical and non-symmetrical, 
for S\'ersic/core-S\'ersic and for early/late-type galaxies.
Symmetrical regressions are meant to be compared with theoretical expectations, 
whereas non-symmetrical forward ($M_{\rm BH}|X$) regressions -- 
which minimize the scatter in the $\log(M_{\rm BH})$ direction -- 
are best used to predict black hole masses.

\section{Results and discussion}
\label{sec:res}

\subsection{Black hole mass -- galaxy luminosity}
\begin{figure}[h]
\begin{center}
\includegraphics[width=\columnwidth]{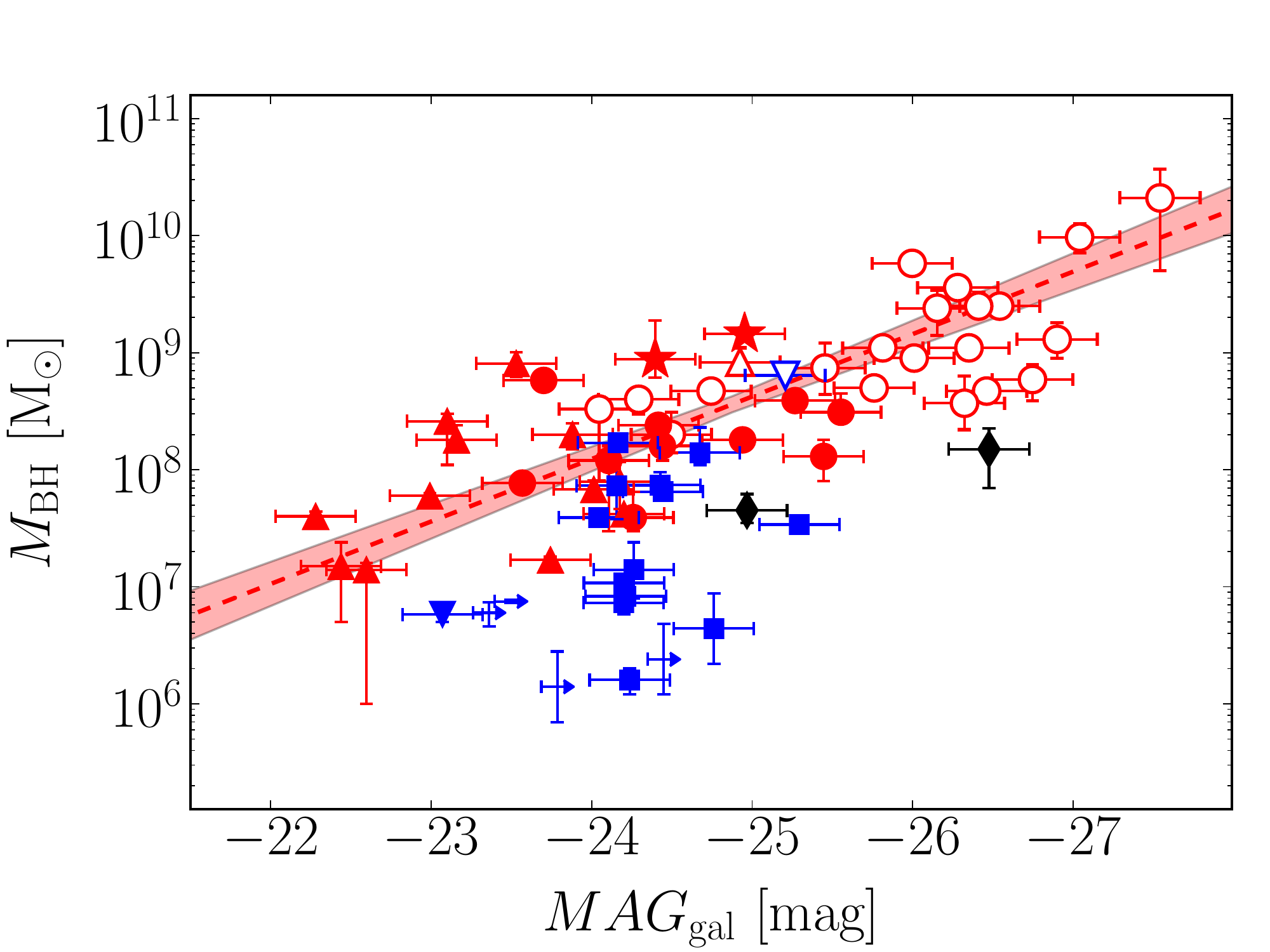}
\caption{Black hole mass plotted against $3.6\rm~\mu m$ galaxy absolute magnitude. 
Symbols are coded according to the galaxy morphological type: red circle = E, red star = E/S0, 
red upward triangle = S0, blue downward triangle = S0/Sp, blue square = Sp, black diamond = merger. 
Empty symbols represent core-S\'ersic spheroids, whereas filled symbols are used for S\'ersic spheroids. 
Four spiral galaxies had their magnitudes overestimated (luminosities underestimated) and are shown as upper limits. 
The red dashed line indicates the BCES bisector linear regression for the 45 early-type galaxies (E+S0), 
with the red shaded area denoting its $1\sigma$ uncertainty. 
$M_{\rm BH}$ correlates equally well with $L_{\rm gal}$ and $L_{\rm sph}$ only for early-type galaxies, 
but not for all (early+late-type) galaxies. }
\label{fig:mbhmaggal}
\end{center}
\end{figure}

The $M_{\rm BH} - L_{\rm gal}$ diagram is shown in Figure \ref{fig:mbhmaggal}.
Four spiral galaxies had their total luminosities underestimated (see Table \ref{tab:sample}) 
and thus are not included in the linear regression analysis (see Table \ref{tab:lreggal}). \\
\cite{lasker2014anal} analyzed a sample of 35 galaxies, among which only four were classified as spiral galaxies, 
and claimed that the $M_{\rm BH} - L_{\rm sph}$ and $M_{\rm BH} - L_{\rm gal}$ relations, 
which they fit with a single power-law, have consistent intrinsic scatter.
Here, instead, thanks to our galaxy sample that includes 17 spiral galaxies, 
we show that the claim made by \cite{lasker2014anal} is valid only for early-type galaxies. 
That is, when considering only early-type galaxies, 
we find that the $M_{\rm BH} - L_{\rm sph}$ and $M_{\rm BH} - L_{\rm gal}$ relations have the same level of intrinsic scatter. 
However, our $M_{\rm BH} - L_{\rm sph}$ relation for all 66 galaxies, irrespective of their morphological type, 
has an intrinsic scatter $\epsilon_{(Y|X)} = 0.51 \pm 0.06\rm~dex$ (forward linear regression) 
and $\epsilon_{(X|Y)} = 0.60 \pm 0.09\rm~dex$ (inverse linear regression), 
whereas our $M_{\rm BH} - L_{\rm gal}$ relation for 62 ($=66-4$) galaxies has $\epsilon_{(Y|X)} = 0.63 \pm 0.07\rm~dex$ 
and $\epsilon_{(X|Y)} = 0.91 \pm 0.17\rm~dex$.
Because the value of the intrinsic scatter depends on the size of the uncertainties associated with the absolute magnitudes\footnote{The 
smaller (larger) the uncertainties, the larger (smaller) the intrinsic scatter.}, 
we tested the robustness of our conclusion by increasing the uncertainties associated with the galaxy absolute magnitudes\footnote{The 
value of the intrinsic scatter obviously depends also on the size of the uncertainties associated with the black hole masses. 
However, black hole masses and their uncertainties have been estimated by various authors using different methods, 
thus we have limited to no control on their values.}  
(we originally assumed $0.25\rm~mag$).
The intrinsic scatter of the $M_{\rm BH} - L_{\rm gal}$ relation only becomes smaller than that of the $M_{\rm BH} - L_{\rm sph}$ relation 
when assuming an uncertainty larger than $0.7\rm~mag$ for $L_{\rm gal}$, 
which would be significantly larger than the typical value commonly recognized in the literature, 
and would also oddly exceed the typical uncertainty that we estimated for $L_{\rm sph}$. 
Hence, we conclude that our determination of the relative intrinsic scatter is reliable, 
and that $M_{\rm BH}$ correlates equally well with $L_{\rm sph}$ and $L_{\rm gal}$ only for early-type galaxies\footnote{The majority 
of our early-type galaxies are elliptical galaxies, some of which have a bulge-to-total ratio close to 1 ($L_{\rm gal} \simeq L_{\rm sph}$).
One might wonder whether this constitutes a bias in our analysis.
However, we checked that $M_{\rm BH}$ correlates equally well with $L_{\rm sph}$ and $L_{\rm gal}$ 
not only for early-type (elliptical + lenticular) galaxies, 
but also for lenticular galaxies only. }, 
but not for all (early+late-type) galaxies. 

\begin{figure}[h]
\begin{center}
\includegraphics[width=\columnwidth]{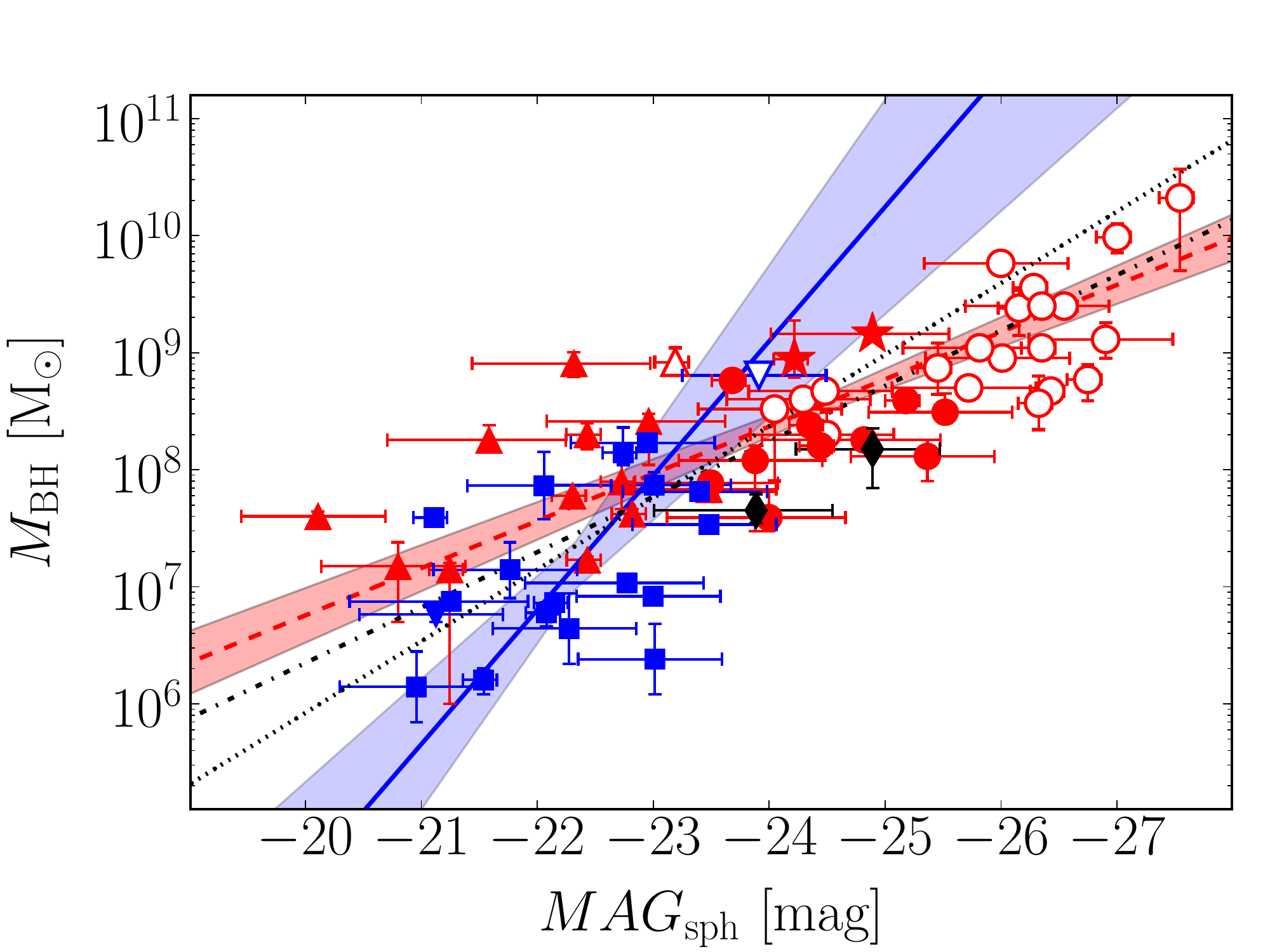}
\caption{Black hole mass plotted against $3.6\rm~\mu m$ spheroid absolute magnitude. 
Symbols have the same meaning as in Figure \ref{fig:mbhmaggal}.
The red dashed line indicates the BCES bisector linear regression for the spheroidal component of the 45 early-type (E+S0) galaxies, 
with the red shaded area denoting its $1\sigma$ uncertainty. 
The blue solid line shows the BCES bisector linear regression for the bulges of the 17 late-type (Sp) galaxies, 
with the blue shaded area denoting its $1\sigma$ uncertainty. 
The black dashed-dotted and dotted lines represent the BCES bisector linear regressions for the core-S\'ersic and S\'ersic spheroids, respectively.}
\label{fig:mbhmagsph}
\end{center}
\end{figure}

\begin{table*}
\centering
\caption{Linear regression analysis of the $M_{\rm BH} - L_{\rm gal}$ diagram.}
\begin{tabular}{llccccc}
\tableline\tableline
{\bf Subsample (size)} & {\bf Regression} & $\boldsymbol \alpha$ & $\boldsymbol \beta$ & $\boldsymbol \langle MAG_{\rm gal} \rangle$ & $\boldsymbol \epsilon$ & $\boldsymbol \Delta$ \\ 
\tableline 
\\
  & \multicolumn{6}{l}{$\log[M_{\rm BH}/{\rm M_\odot}] = \alpha + \beta[(MAG_{\rm gal} - \langle MAG_{\rm gal} \rangle)/{\rm mag}]$} \\ [0.5em]
 All (62)               & BCES $(Y|X)$   & $8.26 \pm 0.08$ & $-0.49 \pm 0.06$ & $-24.78$ & $-$ & $0.64$ \\
                        & mFITEXY $(Y|X)$  & $8.26^{+0.08}_{-0.08}$ & $-0.49^{+0.06}_{-0.07}$ & $-24.78$ & $0.61^{+0.07}_{-0.06}$ & $0.64$ \\
                        & {\tt linmix\_err} $(Y|X)$  & $8.26 \pm 0.09$ & $-0.49 \pm 0.07$ & $-24.78$ & $0.63 \pm 0.07$ & $0.64$ \\ [0.5em]
                        & BCES $(X|Y)$   & $8.26 \pm 0.12$ & $-1.01 \pm 0.15$ & $-24.78$ & $-$ & $0.92$ \\
                        & mFITEXY $(X|Y)$  & $8.26^{+0.11}_{-0.12}$ & $-1.03^{+0.13}_{-0.16}$ & $-24.78$ & $0.88^{+0.10}_{-0.08}$ & $0.93$ \\
                        & {\tt linmix\_err} $(X|Y)$  & $8.26 \pm 0.12$ & $-1.02 \pm 0.15$ & $-24.78$ & $0.91 \pm 0.17$ & $0.93$ \\ [0.5em]
                        & BCES Bisector  & $8.26 \pm 0.09$ & $-0.72 \pm 0.07$ & $-24.78$ & $-$ & $0.71$ \\
                        & mFITEXY Bisector & $8.26^{+0.10}_{-0.10}$ & $-0.73^{+0.09}_{-0.10}$ & $-24.78$ & $-$    & $0.71$ \\
                        & {\tt linmix\_err} Bisector & $8.26 \pm 0.10$ & $-0.72 \pm 0.07$ & $-24.78$ & $-$    & $0.71$ \\ [0.5em]

 Early-type (E+S0) (45) & BCES $(Y|X)$   & $8.56 \pm 0.07$ & $-0.44 \pm 0.05$ & $-24.88$ & $-$ & $0.45$ \\
                        & mFITEXY $(Y|X)$  & $8.56^{+0.06}_{-0.06}$ & $-0.42^{+0.05}_{-0.05}$ & $-24.88$ & $0.41^{+0.06}_{-0.05}$ & $0.45$ \\
                        & {\tt linmix\_err} $(Y|X)$  & $8.56 \pm 0.07$ & $-0.42 \pm 0.06$ & $-24.88$ & $0.43 \pm 0.06$ & $0.45$ \\ [0.5em]
                        & BCES $(X|Y)$   & $8.56 \pm 0.08$ & $-0.64 \pm 0.05$ & $-24.88$ & $-$ & $0.53$ \\
                        & mFITEXY $(X|Y)$  & $8.56^{+0.08}_{-0.08}$ & $-0.66^{+0.07}_{-0.08}$ & $-24.88$ & $0.51^{+0.07}_{-0.06}$ & $0.55$ \\
                        & {\tt linmix\_err} $(X|Y)$  & $8.56 \pm 0.09$ & $-0.65 \pm 0.08$ & $-24.88$ & $0.53 \pm 0.10$ & $0.54$ \\ [0.5em]
                        & BCES Bisector  & $8.56 \pm 0.07$ & $-0.53 \pm 0.04$ & $-24.88$ & $-$ & $0.47$ \\
                        & mFITEXY Bisector & $8.56^{+0.07}_{-0.07}$ & $-0.54^{+0.06}_{-0.06}$ & $-24.88$ & $-$    & $0.47$ \\
                        & {\tt linmix\_err} Bisector & $8.56 \pm 0.08$ & $-0.53 \pm 0.05$ & $-24.88$ & $-$    & $0.47$ \\ [0.5em]

\tableline 
\tableline
\end{tabular}
\label{tab:lreggal} 
\tablecomments{For each subsample, we indicate $\langle MAG_{\rm gal} \rangle$, its average value of galaxy magnitudes. 
In the last two columns, we report $\epsilon$, the intrinsic scatter, and $\Delta$, the total rms scatter in the $\log(M_{\rm BH})$ direction. 
Four spiral galaxies had their luminosities underestimated and thus are not included in the linear regression analysis 
(the sample of all galaxies contains 66-4=62 objects). 
When considering all galaxies, irrespective of their morphological type, 
the $M_{\rm BH} - L_{\rm gal}$ correlation is weaker than the $M_{\rm BH} - L_{\rm sph}$ correlation, in terms of intrinsic scatter. 
However, when considering only early-type galaxies, the $M_{\rm BH} - L_{\rm gal}$ and $M_{\rm BH} - L_{\rm sph}$ correlations 
have consistent intrinsic scatter. }
\end{table*}

\subsection{Black hole mass -- spheroid luminosity}
The $M_{\rm BH} - L_{\rm sph}$ diagram is shown in Figure \ref{fig:mbhmagsph}, 
and the linear regression analysis is presented in Table \ref{tab:lregsph}. 
S\'ersic and core-S\'ersic spheroids have slopes consistent with each other (within their $1\sigma$ uncertainties), 
in disagreement with the findings of GS13. 
The slope that we obtained for core-S\'ersic spheroids ($M_{\rm BH} \propto L_{\rm sph}^{1.18 \pm 0.20}$) 
is consistent with the slope reported by GS13 in the $K_s$-band for the same population ($M_{\rm BH} \propto L_{\rm sph}^{1.10 \pm 0.20}$). 
However, the slope that we determined for S\'ersic spheroids ($M_{\rm BH} \propto L_{\rm sph}^{1.53 \pm 0.20}$) 
is notably shallower than that found by GS13 ($M_{\rm BH} \propto L_{\rm sph}^{2.73 \pm 0.55}$). \\
Although the S\'ersic/core-S\'ersic classification used by GS13 slightly differs\footnote{The classification has changed for the galaxies 
NGC 1316, NGC 1332 and NGC 3998.} from the classification used here, 
the main cause of the inconsistency is that the bulge-to-total ratios obtained from our galaxy decompositions 
are different from those assumed by GS13 to convert galaxy luminosities into bulge luminosities.
Our bulge-to-total ratios for low-luminosity S\'ersic spheroids ($3.6\rm~\mu m$ $MAG_{\rm sph} \gtrsim -22 \rm~mag$) 
are smaller than those used by GS13. 
The host galaxies of such bulges are late-type, spiral galaxies, 
which typically present a complex morphology (bars, double bars, embedded disks, nuclear components, etc).
Our galaxy models account for the extra components, 
while the average bulge-to-total ratios of GS13 were based on less sophisticated S\'ersic-bulge/exponential-disk decompositions 
which overestimated the bulge luminosity.
This results in our bulge magnitudes being on average $\sim$$0.7\rm~mag$ fainter than in GS13, after accounting for the different wavelength of the data.
At the same time, our bulge-to-total ratios for the high-luminosity S\'ersic spheroids ($3.6\rm~\mu m$ $MAG_{\rm sph} \lesssim -24 \rm~mag$) 
are on average larger than those adopted by GS13.
In this regime, the host systems are early-type galaxies that feature intermediate-scale disks\footnote{Intermediate-scale disks are 
disks of stars fully embedded in the spheroidal component of their galaxy. 
They are typical of ``disky'' elliptical galaxies (e.g.~NGC 3377), 
but they can also be found in other types of host galaxies.
They can be considered an intermediate class between nuclear disks, with sizes $\sim$$10-100\rm~pc$, 
and large-scale disks, that encase the bulge and dominate the light at large radii.}. 
Past bulge/disk decompositions failed to correctly identify the extent of such disks and treated them as large-scale disks, 
thus underestimating the bulge luminosity.
The magnitudes that we obtained for such spheroids are on average $\sim$$0.5\rm~mag$ brighter than in GS13. 
These two effects explain the shallower slope that we obtained for the S\'ersic spheroids. \\
The slope that we obtained for S\'ersic spheroids ($1.53 \pm 0.20$) is not consistent with the value of $2.5$ 
expected from $M_{\rm BH} \propto \sigma^5$ and $L_{\rm sph} \propto \sigma^2$. 
In addition, the S\'ersic and core-S\'ersic spheroids appear not to define two distinct $M_{\rm BH} - L_{\rm sph}$ sequences. 
This leads us to investigate substructure in the $M_{\rm BH} - L_{\rm sph}$ diagram for early- and late-type galaxies.
First, we checked that the elliptical and lenticular galaxies, taken separately, have slopes consistent with each other, 
and thus, taken together, they define a single \emph{early-type sequence} in the $M_{\rm BH} - L_{\rm sph}$ diagram. 
We then fit the early-type galaxies with a single log-linear regression, 
and obtained $M_{\rm BH} \propto L_{\rm sph}^{1.00 \pm 0.10}$. 
We did not find any convincing evidence for the change in slope required for consistency 
with the $M_{\rm BH} - \sigma$ and bent $L_{\rm sph} - \sigma$ correlations. 
Because the change in slope should occur at $M_{\rm BH} > 10^{8 \pm 1}\rm~M_\odot$, 
but all the early-type galaxies in our sample have $M_{\rm BH} \gtrsim 10^7\rm~M_\odot$, 
one possible explanation is that we are still not probing enough low black hole masses for this subsample.
An additional possibility is that there is no sharp transition going from $L_{\rm sph} \propto \sigma^2$ at low luminosities 
to $L_{\rm sph} \propto \sigma^5$ at high luminosities. 
Although the knowledge that many ``elliptical'' galaxies actually contain embedded stellar disks dates back  
at least three decades  
\citep{capaccioli1987,carter1987,rixwhite1990,bender1990,scorzabender1990,nieto1991,rixwhite1992,scorzabender1995},
it is mainly thanks to large integral-field-spectrograph surveys of early-type galaxies, such as the ATLAS$^{\rm 3D}$ Project \citep{cappellari2011}, 
that our view has been further advanced 
and it is now commonly accepted that most ``elliptical'' galaxies contain disks. 
Past studies that investigated the $L_{\rm sph} - \sigma$ diagram might have failed to identify and consequently model the disks 
in intermediate-luminosity, early-type galaxies, thus overestimating $L_{\rm sph}$ 
and mistakenly producing a sharp bend in the $L_{\rm sph} - \sigma$ correlation, rather than a continuously curved relation 
(with $L_{\rm sph} \propto \sigma^{3-4}$ at intermediate luminosities). \\
For the bulges of late-type galaxies we obtained $M_{\rm BH} \propto L_{\rm sph}^{2.88 \pm 0.68}$. 
From a cursory inspection of Figure \ref{fig:mbhmagsph}, one might be tempted to doubt the statistical significance of this ``tentative'' \emph{late-type sequence}. 
However, a visual inspection of the plotted data requires one to take into account the error bars 
when judging-by-eye the strength of a correlation. 
Similarly, the Pearson's and Spearman's correlation coefficients are not applicable because 
they do not take into account the error bars on our data.
We have therefore relied on the quantitative regression analysis.

\subsubsection{Pseudo- versus classical bulges}
\begin{figure}[h]
\begin{center}
\includegraphics[width=\columnwidth]{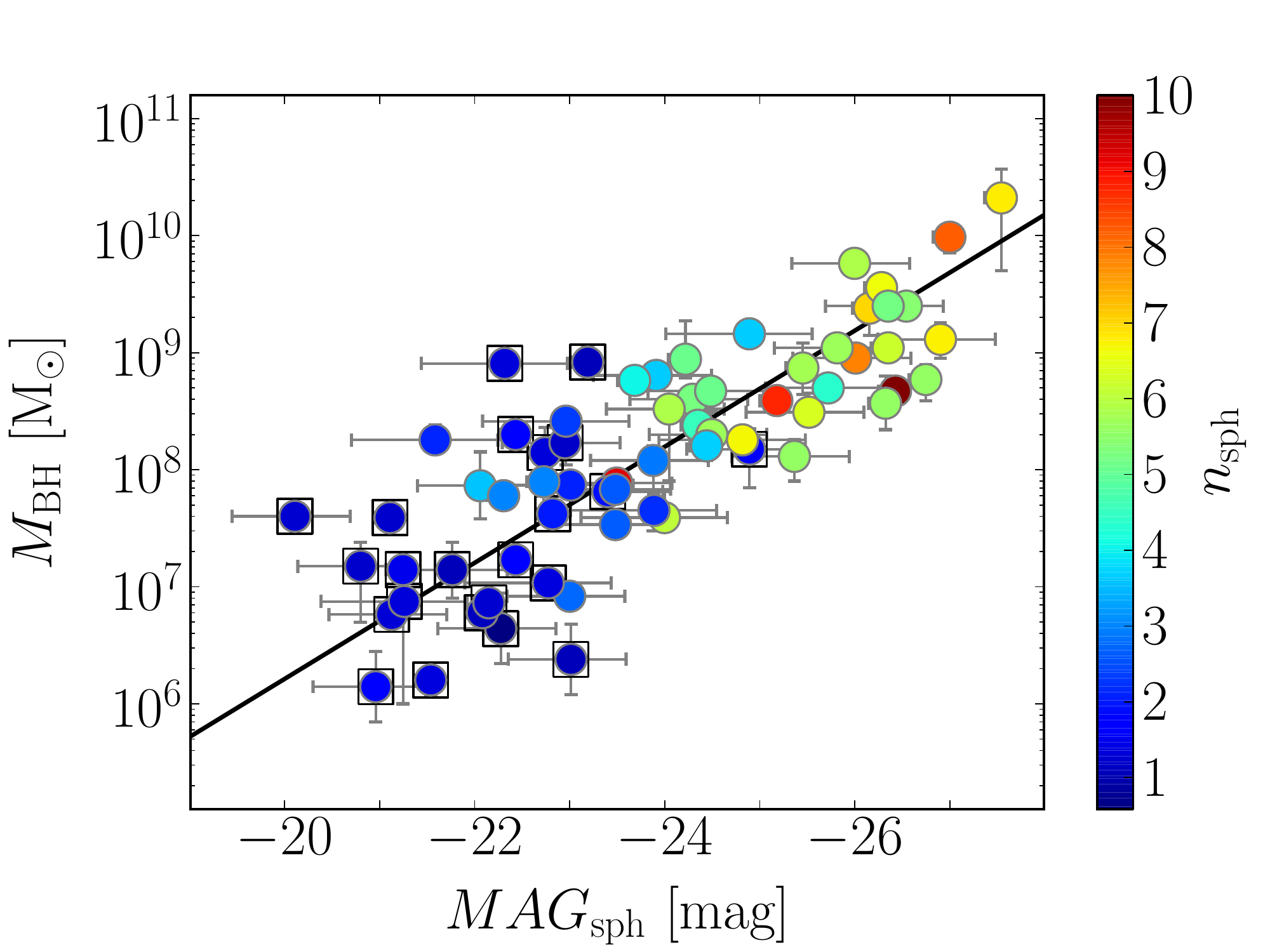}
\caption{Black hole mass plotted against $3.6\rm~\mu m$ spheroid absolute magnitude (as in Figure \ref{fig:mbhmagsph}). 
Symbols are color coded according to the spheroid S\'ersic index $n_{\rm sph}$. 
Bulges with $n_{\rm sph}<2$, claimed by some to be pseudo-bulges, are enclosed with a square. 
The black solid line shows the BCES bisector linear regression for the spheroids that have $n_{\rm sph} \geq 2$, 
such that $M_{\rm BH} \propto L_{\rm sph}^{1.25 \pm 0.13}$. }
\label{fig:pseudob}
\end{center}
\end{figure}

Current views distinguish between classical bulges (which are considered to be spheroidal, pressure-supported systems, 
formed through violent processes, such as hierarchical clustering via minor mergers), 
and pseudo-bulges (thought to be disk-like, rotation-supported systems, 
built from secular evolution processes, such as instabilities of their surrounding disk or bar). 
Pseudo-bulges are notoriously hard to identify \citep{graham2013review,graham2014review,graham2015pseudo,graham2015review}.
For example, mergers can create bulges that rotate (e.g.~\citealt{bekki2010,keselmannusser2012}), 
and bars can spin-up classical bulges (e.g.~\citealt{saha2012,saha2015}), 
thus rotation is not a definitive signature of a pseudo-bulge. 
Furthermore, many galaxies host both a classical and a pseudo-bulge (e.g.~\citealt{erwin2003,erwin2015,
athanassoula2005,gadotti2009,macarthur2009,erwin2010,dosanjosdasilva2013,seidel2015}). 
In the recent literature, pseudo- and classical bulges have frequently been divided at the 
S\'ersic index $n_{\rm sph}=2$ (e.g.~\citealt{sani2011,beifiori2012}), 
although, from a selection of hundreds of disc galaxies imaged in the $K$-band, 
\cite{grahamworley2008} observed no bimodality in the bulge S\'ersic indices about $n_{\rm sph}=2$ or any other value. 
While pseudo-bulges are expected to have exponential-like surface brightness profiles ($n_{\rm sph} \simeq 1$), 
being disky components that formed from their surrounding exponential disks 
(e.g.~\citealt{bardeen1975,hohl1975,combessanders1981,combes1990,pfennigerfriedli1991}), 
it has been shown that mergers can create bulges with $n_{\rm sph}<2$
(e.g.~\citealt{elichemoral2011,scannapieco2011,querejeta2015}), 
just as low-luminosity elliptical galaxies (not built from the secular evolution of a disk)
are also well known to have $n_{\rm sph}<2$ and even $n_{\rm sph}<1$ (e.g.~\citealt{davies1988,youngcurrie1994,jerjen2000}). 
The use of the S\'ersic index (in addition to rotation) to identify pseudo-bulges is thus a dangerous practice. 
We therefore do not assume that all bulges with $n_{\rm sph}<2$ are built from internal processes in the disk 
(i.e.~are what some authors call pseudo-bulges).  \\
\cite{sani2011} reported that pseudo-bulges -- which they labelled as such according to the $n_{\rm sph}<2$ criterion -- 
with low black hole masses ($M_{\rm BH} < 10^7\rm~M_\odot$) are significantly displaced from the correlation 
traced by their (classical) bulges with $n_{\rm sph}>2$. 
In Figure \ref{fig:pseudob}, we show the distribution of spheroid S\'ersic indices\footnote{The spheroid S\'ersic indices 
are taken from our galaxy decompositions (\emph{Paper I}).} 
in the $M_{\rm BH} - L_{\rm sph}$ diagram.
Our aim is to check whether bulges with $n_{\rm sph}<2$ 
are offset to lower black hole masses from the correlation defined by bulges with $n_{\rm sph}>2$. 
To do this, we fit a symmetrical linear regression to the bulges that have $n_{\rm sph}>2$ 
and we then compute the vertical offset of all bulges from this regression. 
In Figure \ref{fig:pseudob}, we plot the vertical offset against $n_{\rm sph}$. 
Among the 23 bulges with $n_{\rm sph}<2$, 12 have a positive vertical offset and 11 have a negative vertical offset. 
\cite{kormendy2015review} provides a list of many pseudo-bulge classification criteria, including the divide at $n_{\rm sph}=2$, 
and cautions that each individual criterion has a failure rate of 0-25\%. 
If this is true, we should have that no less than 75\% of bulges with $n_{\rm sph}<2$ display a negative vertical offset\footnote{One 
reaches the same conclusion when using the vertical offset from the correlation defined by bulges with $n_{\rm sph}>3$ or even $n_{\rm sph}>4$. 
There are 13 and 10 bulges with $n_{\rm sph}<2$ that lie above and below, respectively, the correlation traced by bulges with $n_{\rm sph}>3$. 
Similarly, there are 15 and 8 bulges with $n_{\rm sph}<2$ that lie above and below, respectively, the correlation traced by bulges with $n_{\rm sph}>4$.}. 
What we observe, instead, is that there are the same number of bulges with $n_{\rm sph}<2$ lying above and below 
the correlation defined by bulges with $n_{\rm sph}>2$, 
and that the amplitude of their offset is the same ($\lesssim 1.5\rm~dex$).
That is, within the current dataset, 
bulges with $n_{\rm sph}<2$ do not appear to be offset from the correlation traced by bulges with $n_{\rm sph}>2$. \\

\begin{figure}[h]
\begin{center}
\includegraphics[width=\columnwidth]{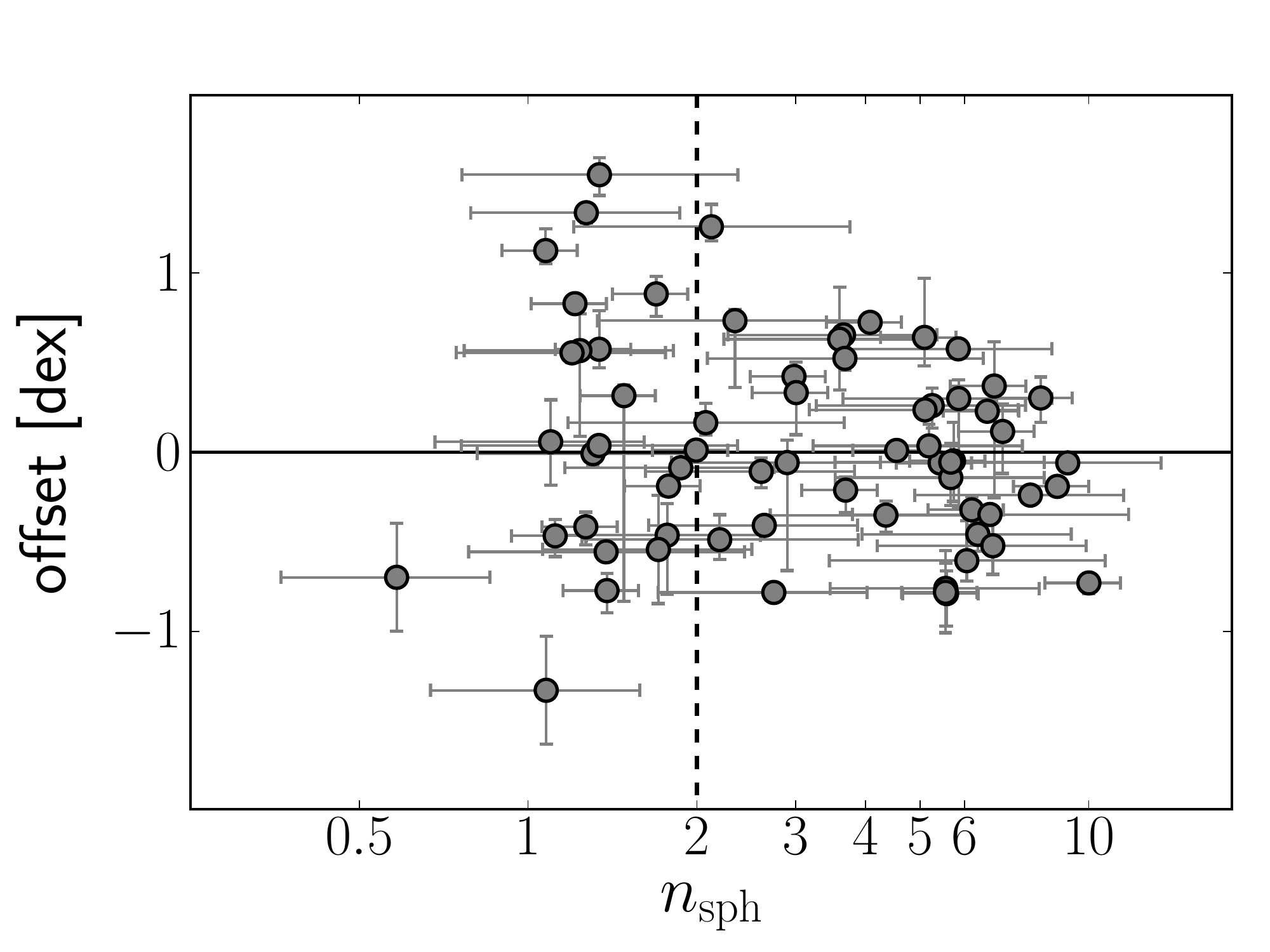}
\caption{Vertical offset from the $M_{\rm BH} - L_{\rm sph}$ correlation defined by spheroids with $n_{\rm sph} \geq 2$ (see Figure \ref{fig:pseudob}), 
plotted against $n_{\rm sph}$. 
The vertical dashed line corresponds to $n_{\rm sph} = 2$.
The horizontal solid line is equivalent to a zero vertical offset.
Among the bulges with $n_{\rm sph}<2$, 12 have a positive vertical offset and 11 have a negative vertical offset.
Hence, bulges with $n_{\rm sph}<2$ are not randomly offset to lower black hole masses 
from the correlation traced by bulges with $n_{\rm sph} \geq 2$.}
\label{fig:offset}
\end{center}
\end{figure}

\begin{table*}
\centering
\caption{Linear regression analysis of the $M_{\rm BH} - L_{\rm sph}$ diagram.}
\begin{tabular}{llccccc}
\tableline
\tableline
{\bf Subsample (size)} & {\bf Regression} & $\boldsymbol \alpha$ & $\boldsymbol \beta$ & $\boldsymbol \langle MAG_{\rm sph} \rangle$ & $\boldsymbol \epsilon$ & $\boldsymbol \Delta$ \\ 
\tableline 
\\
 & \multicolumn{6}{l}{$\log[M_{\rm BH}/{\rm M_\odot}] = \alpha + \beta[(MAG_{\rm sph} - \langle MAG_{\rm sph} \rangle)/{\rm mag}]$} \\ [0.5em]
All (66)               & BCES $(Y|X)$   & $8.16 \pm 0.07$ & $-0.44 \pm 0.04$ & $-23.86$ & $-$ & $0.56$ \\
                       & mFITEXY $(Y|X)$    & $8.17^{+0.06}_{-0.07}$ & $-0.43^{+0.03}_{-0.04}$ & $-23.86$ & $0.49^{+0.06}_{-0.05}$ & $0.56$ \\
                       & {\tt linmix\_err} $(Y|X)$     & $8.16 \pm 0.07$ & $-0.42 \pm 0.04$ & $-23.86$ & $0.51 \pm 0.06$ & $0.56$ \\ [0.5em]
                       & BCES $(X|Y)$   & $8.16 \pm 0.08$ & $-0.61 \pm 0.05$ & $-23.86$ & $-$ & $0.68$ \\
                       & mFITEXY $(X|Y)$    & $8.15^{+0.07}_{-0.08}$ & $-0.61^{+0.05}_{-0.05}$ & $-23.86$ & $0.58^{+0.07}_{-0.06}$ & $0.68$ \\
                       & {\tt linmix\_err} $(X|Y)$     & $8.16 \pm 0.09$ & $-0.60 \pm 0.06$ & $-23.86$ & $0.60 \pm 0.09$ & $0.67$ \\ [0.5em]
                       & BCES Bisector  & $8.16 \pm 0.07$ & $-0.52 \pm 0.04$ & $-23.86$ & $-$ & $0.60$ \\
                       & mFITEXY Bisector   & $8.16^{+0.07}_{-0.07}$ & $-0.51^{+0.04}_{-0.04}$ & $-23.86$ & $-$    & $0.60$ \\
                       & {\tt linmix\_err} Bisector    & $8.16 \pm 0.08$ & $-0.51 \pm 0.03$ & $-23.86$ & $-$    & $0.59$ \\ [0.5em]

$n>2$ (43)             & BCES $(Y|X)$      & $8.58 \pm 0.07$ & $-0.42 \pm 0.06$ & $-24.77$ & $-$ & $0.46$ \\
                       & mFITEXY $(Y|X)$    & $8.57^{+0.07}_{-0.06}$ & $-0.41^{+0.04}_{-0.04}$ & $-24.77$ & $0.38^{+0.06}_{-0.06}$ & $0.46$ \\
                       & {\tt linmix\_err} $(Y|X)$     & $8.56 \pm 0.07$ & $-0.39 \pm 0.05$ & $-24.77$ & $0.40 \pm 0.06$ & $0.46$ \\ [0.5em]
                       & BCES $(X|Y)$      & $8.58 \pm 0.08$ & $-0.58 \pm 0.06$ & $-24.77$ & $-$ & $0.56$ \\
                       & mFITEXY $(X|Y)$    & $8.56^{+0.08}_{-0.08}$ & $-0.57^{+0.06}_{-0.07}$ & $-24.77$ & $0.44^{+0.08}_{-0.11}$ & $0.55$ \\
                       & {\tt linmix\_err} $(X|Y)$     & $8.55 \pm 0.09$ & $-0.57 \pm 0.08$ & $-24.77$ & $0.49 \pm 0.10$ & $0.55$ \\ [0.5em]
                       & BCES Bisector     & $8.58 \pm 0.07$ & $-0.50 \pm 0.05$ & $-24.77$ & $-$ & $0.49$ \\
                       & mFITEXY Bisector   & $8.57^{+0.07}_{-0.07}$ & $-0.49^{+0.05}_{-0.05}$ & $-24.77$ & $-$    & $0.49$ \\
                       & {\tt linmix\_err} Bisector    & $8.56 \pm 0.08$ & $-0.48 \pm 0.05$ & $-24.77$ & $-$    & $0.49$ \\  [0.5em]
                   
Core-S\'ersic (22) & BCES $(Y|X)$   & $9.06 \pm 0.09$ & $-0.32 \pm 0.11$  & $-25.73$ & $-$    & $0.42$ \\
                   & mFITEXY $(Y|X)$   & $9.06^{+0.08}_{-0.09}$ & $-0.26^{+0.08}_{-0.07}$ & $-25.73$ & $0.36^{+0.09}_{-0.06}$ & $0.42$ \\
                   & {\tt linmix\_err} $(Y|X)$  & $9.04 \pm 0.10$ & $-0.24 \pm 0.09$ & $-25.73$ & $0.40 \pm 0.08$ & $0.42$ \\ [0.5em]
                   & BCES $(X|Y)$   & $9.06 \pm 0.12$ & $-0.65 \pm 0.12$  & $-25.73$ & $-$    & $0.61$ \\
                   & mFITEXY $(X|Y)$   & $9.03^{+0.15}_{-0.16}$ & $-0.72^{+0.17}_{-0.31}$ & $-25.73$ & $0.61^{+0.14}_{-0.09}$ & $0.68$ \\
                   & {\tt linmix\_err} $(X|Y)$  & $9.03 \pm 0.17$ & $-0.69 \pm 0.27$ & $-25.73$ & $0.68 \pm 0.30$ & $0.64$ \\ [0.5em]
                   & BCES Bisector  & $9.06 \pm 0.10$ & $-0.47 \pm 0.08$  & $-25.73$ & $-$    & $0.48$ \\
                   & mFITEXY Bisector  & $9.05^{+0.12}_{-0.13}$ & $-0.47^{+0.12}_{-0.17}$ & $-25.73$ & $-$    & $0.48$ \\
                   & {\tt linmix\_err} Bisector & $9.04 \pm 0.14$ & $-0.44 \pm 0.12$ & $-25.73$ & $-$    & $0.46$ \\ [0.5em]

S\'ersic (44) & BCES $(Y|X)$   & $7.71 \pm 0.09$ & $-0.41 \pm 0.08$ & $-22.92$ & $-$    & $0.61$ \\
              & mFITEXY $(Y|X)$   & $7.72^{+0.08}_{-0.09}$ & $-0.41^{+0.07}_{-0.08}$ & $-22.92$ & $0.54^{+0.08}_{-0.07}$ & $0.61$ \\
              & {\tt linmix\_err} $(Y|X)$  & $7.73 \pm 0.09$ & $-0.41 \pm 0.08$ & $-22.92$ & $0.55 \pm 0.08$ & $0.61$ \\ [0.5em]
              & BCES $(X|Y)$   & $7.71 \pm 0.14$ & $-0.86 \pm 0.16$ & $-22.92$ & $-$    & $0.93$ \\
              & mFITEXY $(X|Y)$   & $7.72^{+0.14}_{-0.13}$ & $-0.86^{+0.13}_{-0.19}$ & $-22.92$ & $0.77^{+0.13}_{-0.10}$ & $0.93$ \\
              & {\tt linmix\_err} $(X|Y)$  & $7.73 \pm 0.14$ & $-0.86 \pm 0.17$ & $-22.92$ & $0.79 \pm  0.20$ & $0.93$ \\ [0.5em]
              & BCES Bisector  & $7.71 \pm 0.10$  & $-0.61 \pm 0.08$ & $-22.92$ & $-$    & $0.71$ \\
              & mFITEXY Bisector  & $7.72^{+0.11}_{-0.11}$ & $-0.61^{+0.10}_{-0.12}$ & $-22.92$ & $-$                    & $0.71$ \\
              & {\tt linmix\_err} Bisector & $7.73 \pm 0.12$ & $-0.62 \pm 0.09$ & $-22.92$ & $-$    & $0.71$ \\ [0.5em]

{\bf Early-type (E+S0)} (45) & BCES $(Y|X)$    & $8.56 \pm 0.07$ & $-0.33 \pm 0.04$ & $-24.47$ & $-$    & $0.46$ \\
                             & mFITEXY $(Y|X)$ & $8.56^{+0.06}_{-0.06}$ & $-0.32^{+0.03}_{-0.04}$ & $-24.47$ & $0.40^{+0.06}_{-0.05}$ & $0.46$ \\
                             & {\tt linmix\_err} $(Y|X)$  & $8.55 \pm 0.07$ & $-0.32 \pm 0.04$ & $-24.47$ & $0.41 \pm 0.06$ & $0.46$ \\ [0.5em]
                             & BCES $(X|Y)$    & $8.56 \pm 0.08$ & $-0.48 \pm 0.05$ & $-24.47$ & $-$    & $0.55$ \\
                             & mFITEXY $(X|Y)$ & $8.54^{+0.08}_{-0.08}$ & $-0.49^{+0.05}_{-0.06}$ & $-24.47$ & $0.49^{+0.08}_{-0.06}$ & $0.57$\\
                             & {\tt linmix\_err} $(X|Y)$  & $8.55 \pm 0.09$ & $-0.48 \pm 0.06$ & $-24.47$ & $0.51 \pm 0.10$ & $0.56$ \\ [0.5em]
                             & {\bf BCES Bisector}& $\boldsymbol{8.56 \pm 0.07}$ & $\boldsymbol{-0.40 \pm 0.04}$ & $\boldsymbol{-24.47}$ & $-$    & $\boldsymbol{0.49}$ \\
                             & mFITEXY Bisector   & $8.55^{+0.07}_{-0.07}$ & $-0.41^{+0.04}_{-0.05}$ & $-24.47$ & $-$    & $0.49$ \\
                             & {\tt linmix\_err} Bisector    & $8.55 \pm 0.08$ & $-0.40 \pm 0.04$ & $-24.47$ & $-$    & $0.49$ \\ [0.5em]

{\bf Late-type (Sp)} (17) & BCES $(Y|X)$    & $7.18 \pm 0.16$ & $-0.79 \pm 0.43$ & $-22.33$ & $-$    & $0.70$ \\
                          & mFITEXY $(Y|X)$    & $7.20^{+0.15}_{-0.15}$ & $-0.53^{+0.22}_{-0.24}$ & $-22.33$ & $0.55^{+0.15}_{-0.10}$ & $0.63$ \\
                          & {\tt linmix\_err} $(Y|X)$  & $7.24 \pm 0.19$ & $-0.46 \pm 0.32$ & $-22.33$ & $0.63 \pm 0.16$ & $0.62$ \\ [0.5em]
                          & BCES $(X|Y)$    & $7.18 \pm 0.29$ & $-1.71 \pm 0.71$ & $-22.33$ & $-$    & $1.26$ \\
                          & mFITEXY $(X|Y)$    & $7.38^{+0.54}_{-0.36}$ & $-2.02^{+0.71}_{-2.13}$ & $-22.33$ & $1.09^{+0.41}_{-0.24}$ & $1.50$ \\
                          & {\tt linmix\_err} $(X|Y)$  & $7.34 \pm 0.43$ & $-1.93 \pm 1.30$ & $-22.33$ & $1.31 \pm 0.97$ & $1.43$ \\ [0.5em]
                          & {\bf BCES Bisector}& $\boldsymbol{7.18 \pm 0.20}$ & $\boldsymbol{-1.15 \pm 0.27}$ & $\boldsymbol{-22.33}$ & $-$    & $\boldsymbol{0.88}$ \\
                          & mFITEXY Bisector   & $7.26^{+0.40}_{-0.28}$ & $-1.03^{+0.33}_{-0.52}$ & $-22.33$ & $-$    & $0.82$ \\
                          & {\tt linmix\_err} Bisector & $7.27 \pm 0.33$ & $-0.96 \pm 0.37$ & $-22.33$ & $-$    & $0.78$ \\ [0.5em]

\tableline 
\tableline
\end{tabular}
\label{tab:lregsph} 
\tablecomments{For each subsample, we indicate $\langle MAG_{\rm sph} \rangle$, its average value of spheroid magnitudes. 
In the last two columns, we report $\epsilon$, the intrinsic scatter, and $\Delta$, the total rms scatter in the $\log(M_{\rm BH})$ direction. 
Both the early- and late-type subsamples do not contain the two galaxies classified as S0/Sp and the two galaxies classified as mergers (45+17=66-2-2). }
\end{table*}

\subsection{Black hole mass -- spheroid stellar mass}
\begin{figure}[h]
\begin{center}
\includegraphics[width=\columnwidth]{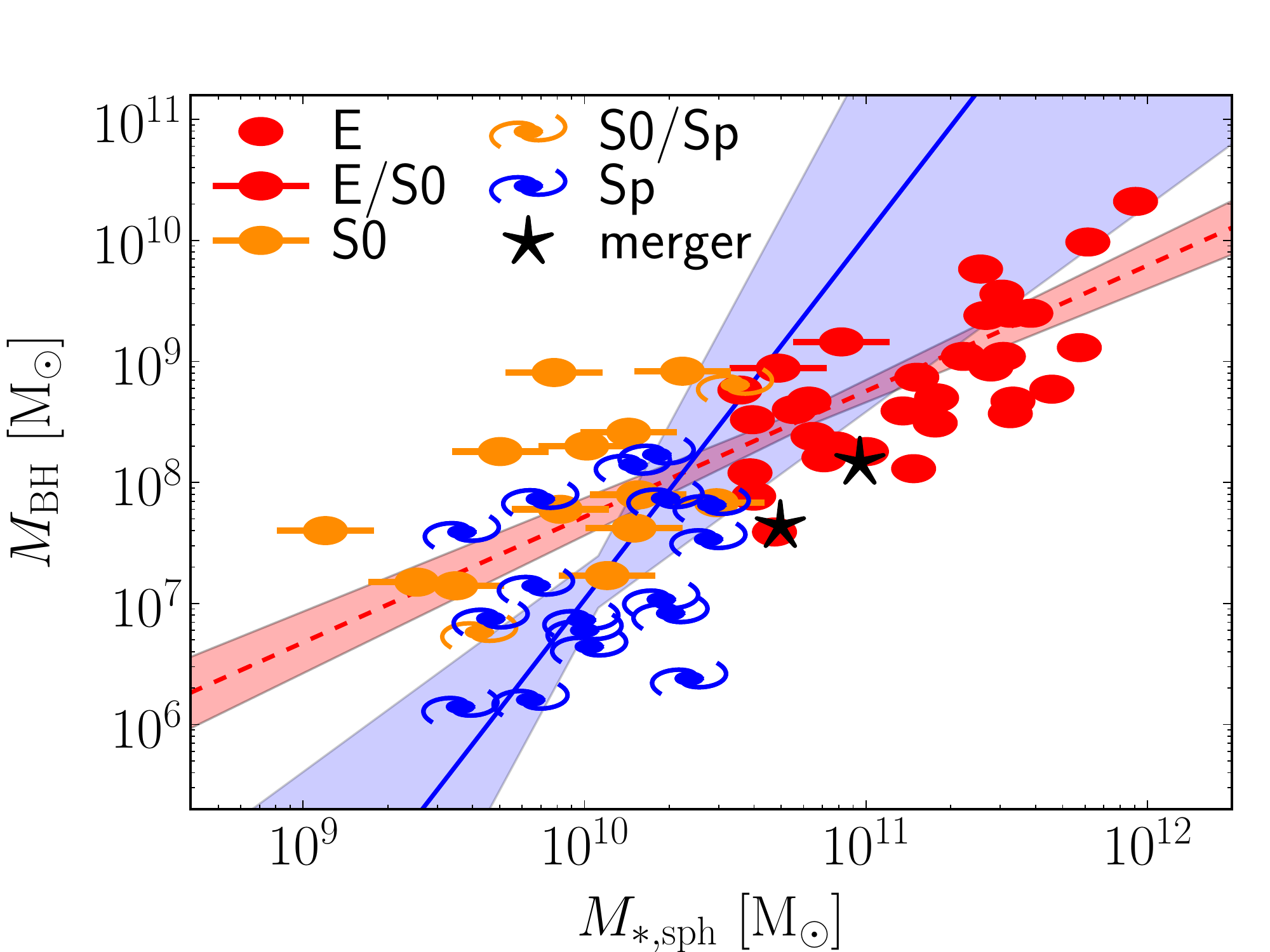}
\caption{Black hole mass plotted against spheroid stellar mass. 
Symbols are coded according to the galaxy morphological type (see legend).
The red dashed line indicates the BCES bisector linear regression for the bulges of the 45 early-type galaxies (E+S0), 
with the red shaded area denoting its $1\sigma$ uncertainty. 
The bulges of early-type galaxies follow $M_{\rm BH} \propto M_{\rm *,sph}^{1.04 \pm 0.10}$,
a near-linear relation consistent with a dry-merging formation scenario.
The steeper blue solid line shows the BCES bisector linear regression for the bulges of the 17 late-type (Sp) galaxies, 
with the blue shaded area denoting its $1\sigma$ uncertainty. 
The bulges of late-type galaxies follow $M_{\rm BH} \propto M_{\rm *,sph}^{2-3}$, 
indicating that gas-rich processes feed the black hole more efficiently (``quadratically'' or ``cubically'') than the host bulge grows in stellar mass. 
We note that AGNs with $10^5 \lesssim M_{\rm BH}/{\rm M_\odot} \lesssim 2 \times 10^6$ \citep{jiang2011a} appear to follow the blue line 
(see Graham et al. 2015, \emph{in preparation}).}
\label{fig:mbhmasssph}
\end{center}
\end{figure}

Finally, we present the $M_{\rm BH} - M_{\rm *,sph}$ diagram in Figure \ref{fig:mbhmasssph}, 
and its linear regression analysis in Table \ref{tab:lregmass}. 
The bulges of the early-type galaxies follow $M_{\rm BH} \propto M_{\rm *,sph}^{1.04 \pm 0.10}$,
consistent with a dry-merging formation scenario\footnote{In dry mergers, the black hole and the bulge grow at the same pace, 
increasing their mass in lock step.},
and define a tight \emph{early-type sequence} with intrinsic scatter $\epsilon_{(Y|X)} = 0.43 \pm 0.06\rm~dex$. 
\cite{graham2012bent} reported that the $M_{\rm BH}/M_{\rm dyn,sph}$ ratio for core-S\'ersic galaxies was 0.36\% 
($M_{\rm dyn,sph}$ is the spheroid dynamical mass), 
and discussed the many implications of this.  
Using a larger data sample, \cite{grahamscott2013} reported that the $M_{\rm BH}/M_{\rm *,sph}$ ratio was 0.49\% for core-S\'ersic galaxies.   
Here we find a median $M_{\rm BH}/M_{\rm *,sph}$ ratio of $0.50 \pm 0.04\%$ for the 22 core-S\'ersic galaxies 
and $0.68 \pm 0.04\%$ for the 45 early-type galaxies.  
Among other things, this higher value (previously reported to be $0.1 - 0.2\%$ for all galaxy types, e.g.~\citealt{marconihunt2003}), 
boosts estimates of the black hole mass function and mass density based on galaxy/spheroid luminosity functions. \\
The bulges of the spiral galaxies trace a steeper \emph{late-type sequence}, 
whose slope is less well constrained due to the smaller size of the subsample and, more importantly, 
to the smaller range in $M_{\rm *,sph}$ that the subsample spans. 
For the bulges of spiral galaxies, the BCES code returns a log-linear relation with a slope $= 3.00 \pm 1.30$, 
while the modified FITEXY routine finds a shallower (but still consistent within the $1\sigma$ uncertainty) 
slope $= 2.28^{+1.67}_{-1.01}$.
More data would be welcome to better constrain the slope of this \emph{late-type sequence}, 
although we note that direct measurements of black hole masses below $10^6\rm~M_\odot$ are extremely challenging to obtain 
with the current technological resources. 
In this regard, using a sample of $\sim$140 low-redshift ($z \leq 0.35$, with a median redshift $\langle z \rangle = 0.085$) 
bulges hosting Active Galactic Nuclei (AGNs) with virial
black hole masses $10^5 \lesssim M_{\rm BH}/{\rm M_\odot} \lesssim 2 \times 10^6$ \citep{jiang2011a}, 
\cite{grahamscott2015} showed that they roughly follow the quadratic $M_{\rm BH} - M_{\rm *,sph}$ relation defined by their S\'ersic bulges.
The majority of our spiral galaxies host an AGN\footnote{According to the nuclear classification reported on NED 
(NASA Extragalactic Database), among our 17 spiral galaxies, at least 12 host a Seyfert AGN and one hosts a LINER AGN.} and
we anticipate here that the correlation traced by our spiral galaxy bulges 
may track the location of these lower mass AGN in the $M_{\rm BH} - M_{\rm *,sph}$ diagram.
That is, the AGNs appear to be the low-mass continuation of our tentative \emph{late-type sequence} shown in Figure \ref{fig:mbhmasssph} 
and this will be explored with more rigour in a forthcoming paper. \\
 
As a final remark, we comment on the work by \cite{reinesvolonteri2015}, 
who investigated the relationship between black hole mass and total galaxy stellar mass, $M_{\rm *,gal}$. 
Their Figure 8 presents the $M_{\rm BH} - M_{\rm *,gal}$ distribution 
for a sample of $\approx 260$ local AGNs with virial black hole masses  
and for $\approx 80$ galaxies with dynamical black hole masses. 
They concluded that the AGN sample and the early-type galaxies with quiescent black holes define two distinct sequences 
in their $M_{\rm BH} - M_{\rm *,gal}$ diagram; 
these two sequences have similar slope, but a normalization factor different by more than one order of magnitude.  
Since we noted that the \cite{jiang2011a} AGN sample follows the steeper $M_{\rm BH} - M_{\rm *,sph}$ correlation 
traced by our spiral galaxy bulges (the majority of which host an AGN), 
it would be interesting to recover the $M_{\rm BH} - M_{\rm *,sph}$ distribution 
also for the AGN sample of \cite{reinesvolonteri2015}. 
However we do note that there is emerging evidence (e.g.~\citealt{busch2015,subramanian2015}) 
for a population of bulges with black hole masses residing below (or to the right of) the red and blue $M_{\rm BH} - M_{\rm *,sph}$ 
sequences constructed here using samples with directly measured black hole masses, as speculated by Batcheldor (2010).

\section{Conclusions}
Using $3.6\rm~\mu m$ \emph{Spitzer} images, 
we have performed accurate multicomponent decompositions (i.e.~bulge, disks, bars, spiral arms, rings, halo, nucleus, depleted core, etc.), 
which were checked to be consistent with the two-dimensional galaxy kinematics, 
for 66 nearby galaxies with a dynamical measurement of their black hole mass.
We have derived galaxy luminosities, spheroid luminosities and spheroid stellar masses. 
Our galaxy sample, besides being to date the largest sample with reliable bulge masses used to investigate black hole mass scaling relations, 
contains 17 spiral galaxies, half of which have $M_{\rm BH} < 10^7 \rm~M_{\odot}$. 
This constitutes a significant improvement over past studies whose samples were biased towards high-mass, early-type galaxies. \\
Using our state-of-the-art dataset, we have investigated substructure in the $M_{\rm BH} - L_{\rm gal}$, $M_{\rm BH} - L_{\rm sph}$ 
and $M_{\rm BH} - M_{\rm *,sph}$ diagrams. 
Our principal conclusions are: \\

\begin{itemize}
\item The logarithmic $M_{\rm BH} - M_{\rm *,sph}$ relation for the spheroidal components of early-type (elliptical + lenticular) galaxies 
      has a slope of $1.04 \pm 0.10$ and intrinsic scatter $\epsilon_{(Y|X)} = 0.43 \pm 0.06\rm~dex$. 
      We call this tight correlation an \emph{early-type sequence}. 
      The $M_{\rm BH} - M_{\rm *,sph}$ log-relation for the bulges of late-type (spiral) galaxies has a slope of $2-3$,
      which is less well constrained due to
      the smaller size of the subsample and, more importantly, the smaller range in spheroid stellar mass 
      ($3 \times 10^9 \lesssim M_{\rm *,sph}/{\rm M_\odot} \lesssim 3 \times 10^{10}$) that the subsample spans. 
      We refer to this correlation as a \emph{late-type sequence}. 
      In (gas-poor) early-type galaxies, the black hole and the stellar content of the spheroidal component grow at the same pace, 
      following a linear $M_{\rm BH} - M_{\rm *,sph}$ relation. 
      In (gas-rich) spiral galaxies, the black hole grows faster than its host bulge, 
      following a quadratic/cubic $M_{\rm BH} - M_{\rm *,sph}$ relation. 
      Unsurprisingly, in a color-magnitude diagram\footnote{Total $B-V$ colors, 
      corrected for inclination, Galactic extinction and K-correction, 
      were taken from the HyperLEDA online database \citep{hyperleda}.}, 
      our early- and late-type galaxies occupy the two distinct regions of the red sequence and the blue cloud, respectively. 
      For analogy with this, we refer to our \emph{early-type sequence} as a \emph{red sequence} 
      and to our \emph{late-type sequence} as a \emph{blue sequence}. 
\item The median $M_{\rm BH}/M_{\rm *,sph}$ ratio for the early-type galaxies is $0.68 \pm 0.04\%$. 
      This value is dramatically larger than what was previously reported ($0.1 - 0.2\%$ for all galaxy types, e.g.~\citealt{marconihunt2003}), 
      but in close agreement with the value of $0.49\%$ reported by \cite{grahamscott2013} for core-S\'ersic spheroids. 
\item The logarithmic $M_{\rm BH} - M_{\rm *,sph}$ relations for the core-S\'ersic and S\'ersic spheroids 
      have slopes with over-lapping uncertainties ($1.19 \pm 0.23$ and $1.48 \pm 0.20$, respectively).  
      The S\'ersic relation is less steep than, but also has over-lapping uncertainties with, 
      the slope of $2.22 \pm 0.58$ reported by \cite{scott2013} for S\'ersic spheroids.  
      The distinction between core-S\'ersic and S\'ersic spheroids found by \cite{scott2013} is thus less pronounced here. 
\item In the $M_{\rm BH} - L_{\rm sph}$ (or $M_{\rm BH} - M_{\rm *,sph}$) diagram, for early-type galaxies, 
      we did not observe the change in slope required for consistency with the log-linear $M_{\rm BH} - \sigma$ and bent $L_{\rm sph} - \sigma$ correlations.
      This issue of inconsistency remains therefore an open question. 
      It might be that we are still not probing enough low-mass black holes ($M_{\rm BH} < 10^7\rm~M_\odot$) 
      for the subsample of early-type galaxies, 
      or that the transition from  $L_{\rm sph} \propto \sigma^2$ at low luminosities to $L_{\rm sph} \propto \sigma^{(5-6)}$ at high luminosities 
      is less sharp than previously thought.
      We intend to investigate this point in our future work. 
\item It has been argued that pseuso-bulges (disk-like, rotation-supported systems, built from secular processes) 
      do not follow the $M_{\rm BH} - L_{\rm sph}$ correlation defined by classical bulges 
      (spheroidal, pressure-supported systems, formed through violent processes).  
      The recent literature (e.g.~\citealt{sani2011,beifiori2012}) has distinguished between pseudo- and classical bulges 
      according to their S\'ersic index, $n_{\rm sph}$. 
      Although we do not consider the S\'ersic index a good indicator of the nature of a bulge (e.g.~\citealt{grahamworley2008}), 
      we investigated this point and found that, within the current dataset,  
      spheroids with $n_{\rm sph}<2$ 
      are not offset to lower $M_{\rm BH}$ from the $M_{\rm BH} - L_{\rm sph}$ correlation defined by spheroids with $n_{\rm sph}>2$. 
\item The $M_{\rm BH} - L_{\rm gal}$ and $M_{\rm BH} - L_{\rm sph}$ correlations have the same level of intrinsic scatter 
      when considering early-type galaxies only. 
      Once reasonable numbers of spiral galaxies are included, 
      $M_{\rm BH}$ correlates better with $L_{\rm sph}$ than with $L_{\rm gal}$ (see also \citealt{erwingadotti2012,beifiori2012}). 
\end{itemize} 

Finally, we note that some of the literature-sourced black hole mass measurements used by \cite{kormendyho2013} are different from those used here.  
While these differences are smaller than $18\%$ for $78\%$ of the galaxies, 
in three cases (NGC 0821, NGC 4291, and NGC 3393) they are larger than a factor of $2.3$.  
We repeated our entire analysis using only the 58 galaxies that are in common between our sample and the sample of \cite{kormendyho2013}, 
assuming for these galaxies the black hole mass measurements published by \cite{kormendyho2013}. 
In doing so, we found that none of our conclusions changed.

\begin{table*}
\centering
\caption{Linear regression analysis of the $M_{\rm BH} - M_{\rm *,sph}$ diagram.}
\begin{tabular}{llccccc}
\tableline
\tableline
{\bf Subsample (size)} & {\bf Regression} & $\boldsymbol \alpha$ & $\boldsymbol \beta$ & $\boldsymbol \langle M_{\rm *,sph} \rangle$ & $\boldsymbol \epsilon$ & $\boldsymbol \Delta$ \\ 
\tableline 
\\
  & \multicolumn{6}{l}{$\log[M_{\rm BH}/{\rm M_\odot}] = \alpha + \beta \log[(M_{\rm *,sph} / \langle M_{\rm *,sph} \rangle)]$} \\ [0.5em]
 Core-S\'ersic (22)     & BCES $(Y|X)$      & $9.06 \pm 0.09$ & $0.86 \pm 0.28$ & $10^{11.28}$ & $-$ & $0.42$ \\
 			& mFITEXY $(Y|X)$   & $9.06^{+0.08}_{-0.08}$ & $0.68^{+0.21}_{-0.20}$ & $10^{11.28}$ & $0.36^{+0.09}_{-0.06}$ & $0.42$ \\
  			& {\tt linmix\_err} $(Y|X)$     & $9.04 \pm 0.10$ & $0.64 \pm 0.25$ & $10^{11.28}$ & $0.40 \pm 0.09$ & $0.42$ \\ [0.5em]
			& BCES $(X|Y)$      & $9.06 \pm 0.12$ & $1.70 \pm 0.32$ & $10^{11.28}$ & $-$ & $0.61$ \\
 			& mFITEXY $(X|Y)$   & $9.03^{+0.15}_{-0.16}$ & $1.90^{+0.85}_{-0.46}$ & $10^{11.28}$ & $0.62^{+0.13}_{-0.10}$ & $0.68$ \\
 			& {\tt linmix\_err} $(X|Y)$     & $9.03 \pm 0.17$ & $1.80 \pm 0.70$ & $10^{11.28}$ & $0.67 \pm 0.30$ & $0.65$ \\ [0.5em]
 			& BCES Bisector     & $9.06 \pm 0.10$ & $1.19 \pm 0.23$ & $10^{11.28}$ & $-$ & $0.47$ \\
 			& mFITEXY Bisector  & $9.05^{+0.12}_{-0.13}$ & $1.12^{+0.35}_{-0.27}$ & $10^{11.28}$ & $-$                    & $0.46$ \\
 			& {\tt linmix\_err} Bisector	& $9.04 \pm 0.14$ & $1.06 \pm 0.26$ & $10^{11.28}$ & $-$	& $0.45$ \\ [0.5em]

 S\'ersic (44)		& BCES $(Y|X)$      & $7.71 \pm 0.09$ & $0.95 \pm 0.21$ & $10^{10.25}$ & $-$ & $0.64$ \\
 			& mFITEXY $(Y|X)$   & $7.72^{+0.10}_{-0.09}$ & $0.96^{+0.21}_{-0.21}$ & $10^{10.25}$ & $0.58^{+0.09}_{-0.07}$ & $0.64$ \\
 			& {\tt linmix\_err} $(Y|X)$     & $7.73 \pm 0.10$ & $0.98 \pm 0.24$ & $10^{10.25}$ & $0.59 \pm 0.08$ & $0.65$ \\ [0.5em]
 			& BCES $(X|Y)$      & $7.71 \pm 0.16$ & $2.52 \pm 0.54$ & $10^{10.25}$ & $-$ & $1.11$ \\
 			& mFITEXY $(X|Y)$   & $7.72^{+0.16}_{-0.16}$ & $2.49^{+0.69}_{-0.45}$ & $10^{10.25}$ & $0.93^{+0.15}_{-0.13}$ & $1.10$ \\
 			& {\tt linmix\_err} $(X|Y)$     & $7.73 \pm 0.17$ & $2.48 \pm 0.59$ & $10^{10.25}$ & $0.95 \pm 0.27$ & $1.10$ \\ [0.5em]
 			& BCES Bisector     & $7.71 \pm 0.11$ & $1.48 \pm 0.20$ & $10^{10.25}$ & $-$ & $0.74$ \\
 			& mFITEXY Bisector  & $7.72^{+0.13}_{-0.13}$ & $1.49^{+0.33}_{-0.28}$ & $10^{10.25}$ & $-$    & $0.74$ \\
 			& {\tt linmix\_err} Bisector	& $7.73 \pm 0.14$ & $1.49 \pm 0.24$ & $10^{10.25}$ & $-$	& $0.74$ \\ [0.5em]

{\bf Early-type (E+S0)} (45)  & BCES $(Y|X)$       & $8.56 \pm 0.07$ & $0.85 \pm 0.12$ & $10^{10.81}$ & $-$ & $0.48$ \\
                              & mFITEXY $(Y|X)$     & $8.56^{+0.06}_{-0.07}$ & $0.83^{+0.11}_{-0.11}$ & $10^{10.81}$ & $0.42^{+0.07}_{-0.05}$ & $0.48$ \\
                              & {\tt linmix\_err} $(Y|X)$     & $8.55 \pm 0.07$ & $0.82 \pm 0.12$ & $10^{10.81}$ & $0.43 \pm 0.06$ & $0.48$ \\ [0.5em]
                              & BCES $(X|Y)$       & $8.56 \pm 0.09$ & $1.27 \pm 0.13$ & $10^{10.81}$ & $-$ & $0.59$ \\
                              & mFITEXY $(X|Y)$     & $8.54^{+0.08}_{-0.09}$ & $1.32^{+0.18}_{-0.15}$ & $10^{10.81}$ & $0.53^{+0.08}_{-0.07}$ & $0.61$ \\
                              & {\tt linmix\_err} $(X|Y)$     & $8.55 \pm 0.09$ & $1.29 \pm 0.17$ & $10^{10.81}$ & $0.54 \pm 0.11$ & $0.59$ \\ [0.5em]
                              & {\bf BCES Bisector}& $\boldsymbol{8.56 \pm 0.07}$ & $\boldsymbol{1.04 \pm 0.10}$ & $\boldsymbol{10^{10.81}}$ & $-$ & $\boldsymbol{0.51}$ \\
                              & mFITEXY Bisector    & $8.55^{+0.07}_{-0.08}$ & $1.05^{+0.14}_{-0.12}$ & $10^{10.81}$ & $-$                    & $0.51$ \\
                              & {\tt linmix\_err} Bisector    & $8.55 \pm 0.08$ & $1.03 \pm 0.10$ & $10^{10.81}$ & $-$    & $0.51$ \\ [0.5em]

{\bf Late-type (Sp)} (17)    & BCES $(Y|X)$    & $7.18 \pm 0.17$ & $1.95 \pm 1.52$ & $10^{10.05}$ & $-$ & $0.74$ \\ 
                             & mFITEXY $(Y|X)$     & $7.20^{+0.15}_{-0.16}$ & $1.22^{+0.70}_{-0.62}$  & $10^{10.05}$ & $0.59^{+0.16}_{-0.11}$ & $0.66$ \\
                             & {\tt linmix\_err} $(Y|X)$  & $7.23 \pm 0.19$ & $0.96 \pm 0.96$ & $10^{10.05}$ & $0.67 \pm 0.16$ & $0.65$ \\ [0.5em]
                             & BCES $(X|Y)$    & $7.18 \pm 0.39$ & $5.89 \pm 3.40$ & $10^{10.05}$ & $-$ & $1.70$ \\
                             & mFITEXY $(X|Y)$     & $7.44^{+1.45}_{-0.52}$ & $7.14^{+26.31}_{-3.01}$ & $10^{10.05}$ & $1.49^{+0.56}_{-0.36}$ & $2.08$ \\
                             & {\tt linmix\_err} $(X|Y)$  & $7.42 \pm 0.64$ & $6.96 \pm 6.73$ & $10^{10.05}$ & $1.83 \pm 1.86$ & $2.03$ \\ [0.5em]
                             & {\bf BCES Bisector}& $\boldsymbol{7.18 \pm 0.21}$ & $\boldsymbol{3.00 \pm 1.30}$ & $\boldsymbol{10^{10.05}}$ & $-$ & $\boldsymbol{0.94}$ \\
                             & {\bf mFITEXY Bisector}    & $\boldsymbol{7.24^{+1.04}_{-0.39}}$ & $\boldsymbol{2.28^{+1.67}_{-1.01}}$  & $\boldsymbol{10^{10.05}}$ & $-$    & $\boldsymbol{0.79}$ \\
                             & {\tt linmix\_err} Bisector & $7.26 \pm 0.47$ & $1.94 \pm 1.24$ & $10^{10.05}$ & $-$    & $0.74$ \\ [0.5em]
                  
\tableline 
\tableline
\end{tabular}
\label{tab:lregmass} 
\tablecomments{For each subsample, we indicate $\langle M_{\rm *,sph} \rangle$, its average value of spheroid stellar masses. 
In the last two columns, we report $\epsilon$, the intrinsic scatter, and $\Delta$, the total rms scatter in the $\log(M_{\rm BH})$ direction. }
\end{table*}

\acknowledgments
GS warmly thanks Luca Cortese, Elisabete Lima Da Cunha, Duncan Forbes and Gonzalo Diaz for useful discussion. 
We also thank the anonymous referee for useful comments and suggestions. 
This research was supported by Australian Research Council funding through grants
DP110103509 and FT110100263.
This work is based on observations made with the IRAC instrument \citep{fazio2004IRAC} 
on-board the Spitzer Space Telescope, 
which is operated by the Jet Propulsion Laboratory, 
California Institute of Technology under a contract with NASA.
This research has made use of the GOLDMine database \citep{goldmine} and the NASA/IPAC Extragalactic Database (NED) 
which is operated by the Jet Propulsion Laboratory, California Institute of Technology, 
under contract with the National Aeronautics and Space Administration. 
We acknowledge the usage of the HyperLeda database (\url{http://leda.univ-lyon1.fr}).
The BCES routine \citep{akritasbershady1996} was run via the python module 
written by Rodrigo Nemmen \citep{nemmen2012}, which is available at \url{https://github.com/rsnemmen/BCES}.

\bibliography{SMBHbibliography}

\clearpage

\end{document}